\documentclass[10pt,conference]{IEEEtran} 
\IEEEoverridecommandlockouts
 \pdfoutput=1 
\usepackage{cite}
\usepackage{amsmath,amssymb,amsfonts}
\usepackage{algorithmic}
\usepackage{textcomp}
\usepackage{xcolor}
\def\BibTeX{{\rm B\kern-.05em{\sc i\kern-.025em b}\kern-.08em
    T\kern-.1667em\lower.7ex\hbox{E}\kern-.125emX}}

\usepackage{multirow}
\usepackage{multicol}
\usepackage{booktabs,tabularx}
\newcolumntype{M}[1]{>{\raggedright}m{#1}}

\usepackage{graphicx,latexsym,amsfonts,amsmath,amsthm,amssymb,verbatim,fancyhdr,multicol,babel,subfigure}

\usepackage{float}
\usepackage{graphicx}
\usepackage{braket}
\usepackage{amsfonts}
\begin{document}

\title{Fight or Flight:\\Cosmic Ray-Induced Phonons and the Quantum Surface Code*\\
\thanks{This work is supported by JST Moonshot R$\&$D Grant (JPMJMS2061).}
}

\author{\IEEEauthorblockN{Bernard Ousmane Sane}
\IEEEauthorblockA{\textit{Graduate School of Media and } \\
\textit{Governance, Keio University, 
Japan} \\
bernard@sfc.wide.ad.jp}
\and
\IEEEauthorblockN{Rodney Van Meter}
\IEEEauthorblockA{\textit{Graduate School of Media and } \\
\textit{Governance, Keio University, Japan} \\
rdv@sfc.wide.ad.jp}
\and

\IEEEauthorblockN{Michal Hajdušek}
\IEEEauthorblockA{\textit{Graduate School of Media and}\\
\textit{Governance, Keio University, Japan} \\
michal@sfc.wide.ad.jp}
}

\maketitle

\begin{abstract}
Recent work has identified cosmic ray events as an error source limiting the lifetime of quantum data. These errors are correlated and affect a large number of qubits, leading to the loss of data across a quantum chip. Previous works attempting to address the problem in hardware or by building distributed systems still have limitations. We approach the problem from a different perspective, developing a new hybrid hardware-software-based strategy based on the 2-D surface code, assuming the parallel development of a hardware strategy that limits the phonon propagation radius. We propose to flee the area: move the logical qubits far enough away from the strike's epicenter to maintain our logical information. Specifically, we: (1) establish the minimum hardware requirements needed for our approach; (2) propose a mapping for moving logical qubits; and (3) evaluate the possible choice of the code distance. Our analysis considers two possible cosmic ray events: those far from both ``holes'' in the surface code and those near or overlapping a hole. We show that the probability that the logical qubit will be destroyed can be reduced from 100\% to the range 4\% to 15\% depending on the time required to move the logical qubit.
\end{abstract}


\section{\label{sec:level1}Introduction}

Quantum computers promise a change in the range of problems that can be solved via digital computation but face the crucial challenge of error handling. Error correction can be achieved through coding theory, building a logical qubit which is a grouping of several (or many) physical qubits to reduce the probability of errors. These error correction techniques have been proven to work in quantum systems where errors are uncorrelated \cite{lidar_brun_2013, nielsen_chuang_2010, gottesman2010introduction, devitt13:rpp-qec, RevModPhys.87.307}. Despite this, many worry about cosmic rays impinging on quantum chips since a cosmic ray event (CRE) can produce phonons that induce correlated errors that affect multiple qubits at the same time~\cite{mcewen2021resolving, Wilen_2021, Cardani-paper}. 
A phonon is a quasiparticle, an emergent phenomenon that behaves like an independent quantum particle but isn't a fundamental one. A phonon is a material's quantum vibration unit (usually a crystalline lattice). We don't see them at room temperature because everything constantly vibrates, but phonons can be necessary for several critical quantum technologies at the millikelvin operating temperatures. In superconducting wires, phonons can break apart the pairs of electrons known as Cooper pairs that are central to the operation of superconducting qubits~\cite{Zwanenburgspinqubits, 2012majorana, martiniss415342021saving}.
Quantum error correction would be hindered by these correlated errors~\cite {Clemens2004quantum, 
 Nickerson2019analysingcorrelated}. With important computations on fault-tolerant quantum computers expected to take days or even months~\cite{van-meter13:_blueprint,Gidney2021howtofactorbit}, CREs occurring at the rate of several per minute will impose an unacceptable upper bound on the length of computations that can be successfully executed.

Transmon superconducting qubits use two slightly different energy levels~\cite{PhysRevA.76.042319, krantz2019quantum}. By convention, the $|1\rangle$ state is the higher energy state, and the $|0\rangle$ is the ground state. In this scenario, the cosmic ray strike causes a lot of phonon vibrations in the chip substrate, causing $|1\rangle$ to decay to $|0\rangle$, but not $|0\rangle$ to excite to $|1\rangle$. This asymmetry is one of the signatures that error detection can use to detect strikes. McEwen \emph{et al.} \cite{mcewen2021resolving} set up the system in a state of all ones and looked for correlated decays to zeroes. Hence, they show how the errors spread from the strike location and the damage they can cause across the chip. They determined that in their system, the effects take $25$ milliseconds or so to fade away, a very long time compared to solid-state qubit lifetimes.

As a solution, the authors in \cite{PhysRevLett.129.240502} (including two of the authors of this paper) developed a distributed quantum error correction scheme with two levels of encoding (an intra-chip surface code concatenated with an inter-chip CSS code such as the Steane code). They showed that their proposal reduces the rate of data loss from CREs from the physical event rate of once every $10$ seconds to $1$ loss per month. Suzuki \emph{et al.} propose a fault-tolerant architecture based on superconducting and surface code, in which errors are avoided by dynamically increasing the code distance~\cite{Suzukiburst2022}.

The effect of cosmic rays can also be reduced through engineering approaches that involve changes in hardware~\cite{patel2017phonon, nsanzineza2014trapping, henriques2019phonon}. Through changes on the material level, researchers aim to minimize the surface impact of the cosmic ray by trapping quasiparticles or channeling the cosmic ray's energy ~\cite{patel2017phonon, nsanzineza2014trapping, henriques2019phonon}. To tackle CREs, most of the mitigation strategies try to encompass one or more of the following goals:
\begin{enumerate}
\item Reduce the incidence of strikes,
\item Reduce the range or rate of propagation,
\item Reduce the impact on logical states when a strike happens. 
\end{enumerate}
\begin{figure}[t]
    \centering
    \includegraphics[scale=0.32]{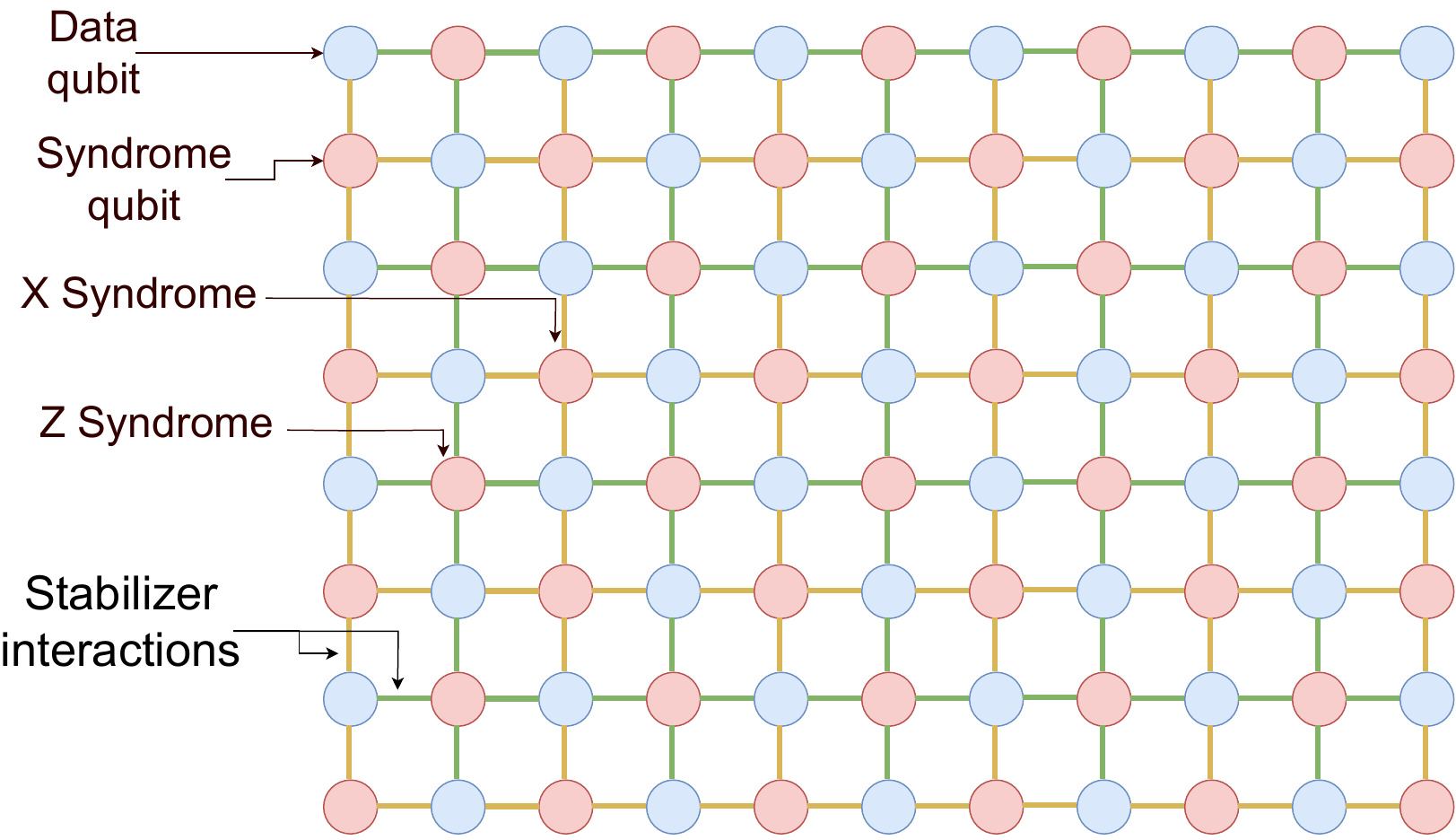}
    \caption{In the 2-D surface code, qubits are coupled only to their neighbors. Half of the qubits (blue) hold an entangled data state. One-quarter of the qubits are used for X stabilizer measurements and the remaining one-quarter for Z stabilizer measurements (red).}
    \label{fig:surface-code}
\end{figure}
However, these approaches still have their limitations. In silicon, sound propagates anisotropically, but we can use 2.5km/sec., or 2.5mm/$\mu$sec, as a reasonable value \cite{mcewen2021resolving}. The measured lifetime of phonons in superconducting chips corresponds to propagation distances greater than $60$ meters before the phonons dissipate. This distance is unlimited compared to the size of superconducting chips. Unfortunately, quantum error correcting codes use physical qubits that are physically close together, presenting a challenge. Thus, we must begin by assuming hardware improvements and then ask how software can contribute to the solution. The first approach above involves reducing the cross section or adding shielding if ambient radioactivity is involved to minimize the strike probability. The second approach is to change the structure of the chip substrate, attempting to dampen the vibrations and reduce their propagation. The third approach involves correcting errors so rapidly that they don't propagate or encode states in long-range correlations so that local destruction isn't a problem.

This paper focuses on software-based strategies for mitigating the effects of cosmic ray hits on systems using the 2-D surface code. Our proposed approach is to flee the area: move logical qubits far enough from the strike's epicenter to preserve our logical information. Hence, the hybrid solution we propose begins with the hardware strategy, which limits the radius of phonon propagation, and ends with software strategy-based surface code. 

\subsection*{Contributions}
Our main contributions are:
\begin{itemize}
    \item The use of 2-D array surface codes in mapping against cosmic ray impacts: we map the qubits so that whenever a cosmic ray occurs, we can minimize the time steps that affect the moving of potentially vulnerable qubits.
    \item To establish goals for the hardware parameters needed to use our approach
\end{itemize}
We know that a fast-moving state depends on the position of the escapement (the nearest open trajectory) and the density of our mapping. Hence, we propose this new hybrid hardware-software-based strategy based on 2-D surface code.
\begin{figure}[t]
    \centering
    \includegraphics[scale=0.5]{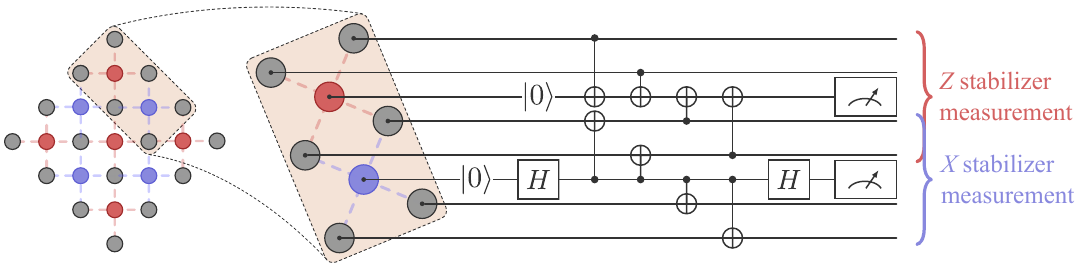}
    \caption{The circuits that perform the surface code Z stabilizer measurement and the X stabilizer measurement can be partially overlapped in execution.}
    \label{fig:measure-syndrome}
\end{figure}
The manuscript is organized as follows:
Surface code is presented in Section \ref{sec:surface-code}. Then, we describe the hardware and software strategies to fight against the cosmic ray in Section \ref{sec:fight}. We present the proposed strategy to flee the strike’s epicenter in Section \ref{sec:fly}. We describe our solution for multiple logical qubits in section \ref{sec:multiqubit}. We analyze our proposals in section \ref{sec:eval}. We conclude this paper with a discussion in Section \ref{sec:discussions-limitations}.

\section{Surface code} 
\label{sec:surface-code}

A surface code is a topological code in which syndromes are measured locally, and fault tolerance can be achieved. The qubits are arranged on a lattice to facilitate interactions between neighboring qubits as illustrated in Fig.~\ref{fig:surface-code}. The 2-D surface code encodes logical qubits in the relationship between boundaries on a specialized, 2-D lattice cluster state~\cite{raussendorf:PhysRevLett.98.190504, fowler:PhysRevA.86.032324, devitt13:rpp-qec}. This relationship can take several forms: a single, independent block (in which logical gates can be executed either transversally or using lattice surgery~\cite{horsman2012surface-njp}), a block with a single ``hole'' cut into it (made by simply not measuring the stabilizers inside of the area, creating a new boundary for the surface), or by using pairs of holes, making for a flexible arrangement and allowing many qubits on a single, large surface. This paper focuses on the Raussendorf two-hole form \cite{RaussendorfPhysRevLett.98.190504}.

Roughly half of the qubits are data qubits, and half are syndrome qubits; stabilizers for groups of four qubits (or three along the edges) are all measured simultaneously for a set of X stabilizers and then for a set of Z stabilizers, as shown in Fig.~\ref{fig:measure-syndrome}. Lattice cycle time $t_c$ refers to the time it takes to measure the $X$ and $Z$ stabilizers and correct any detected errors. Across the appropriate region of the lattice, $N$ data qubits will have $N-1$ stabilizers, leaving the single degree of freedom that becomes our logical qubit. Because the state is encoded in the relationship between two boundaries, a single-qubit logical gate is executed by flipping a string of data qubits connecting the two boundaries. Consequently, logical errors also involve flipping such a string; the code distance $d$ is the number of data qubits in the shortest such string. Because of the structure of the stabilizer measurements and the possibility of errors in stabilizer measurement, this code distance extends not only in the spatial dimension but also in the temporal one, creating a 3-D space-time structure of rectilinear prisms.

\begin{figure}[t]
    \centering
    \includegraphics[scale=0.25]{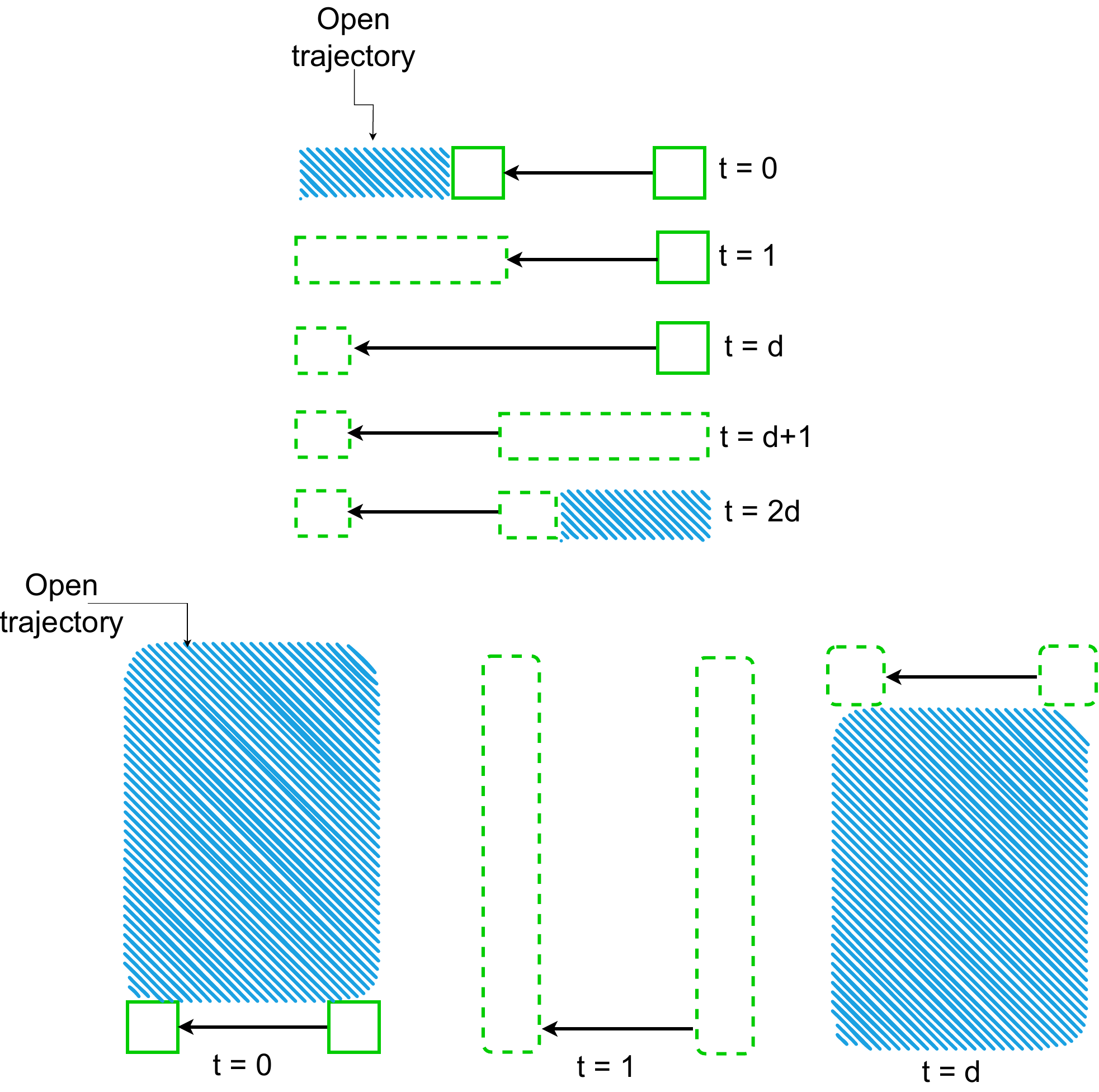}
    \caption{A hole can flee vertically or horizontally any lattice distance along an available open trajectory (blue diagonally shaded areas) in a fixed amount of time (determined by the code distance $d$). Suppose the two holes forming a logical qubit are laid out horizontally. In that case, horizontal movement (above) requires twice as long as vertical movement (below) because one hole blocks the movement of the other, preventing both holes from moving simultaneously. The arrows indicate the inter-hole relationship that expresses the logical qubit.}
   \label{fig:mv-workflow}
\end{figure}

A hole can be moved from one place to another simply by changing the set of stabilizers on the surface measured in each cycle, extending and shrinking the hole, as in Fig.~\ref{fig:mv-workflow}. Two-qubit gates (e.g., CNOT) are executed by \emph{braiding} two pairs of holes by moving the holes on the surface. The time required for this braiding operation depends on the code distance $d$.

\section{Fight: Distance and Delocalization}
\label{sec:fight}

\subsection{Hardware Strategies}

Hybrid strategies must be able to physically limit the maximum radius of phonon propagation (which we will call $r_{\textrm{max}}$), or any software strategy will inevitably be overcome. Hardware techniques come with tradeoffs we must evaluate.

\begin{figure}[H]
    \centerline{\includegraphics[scale=0.22]{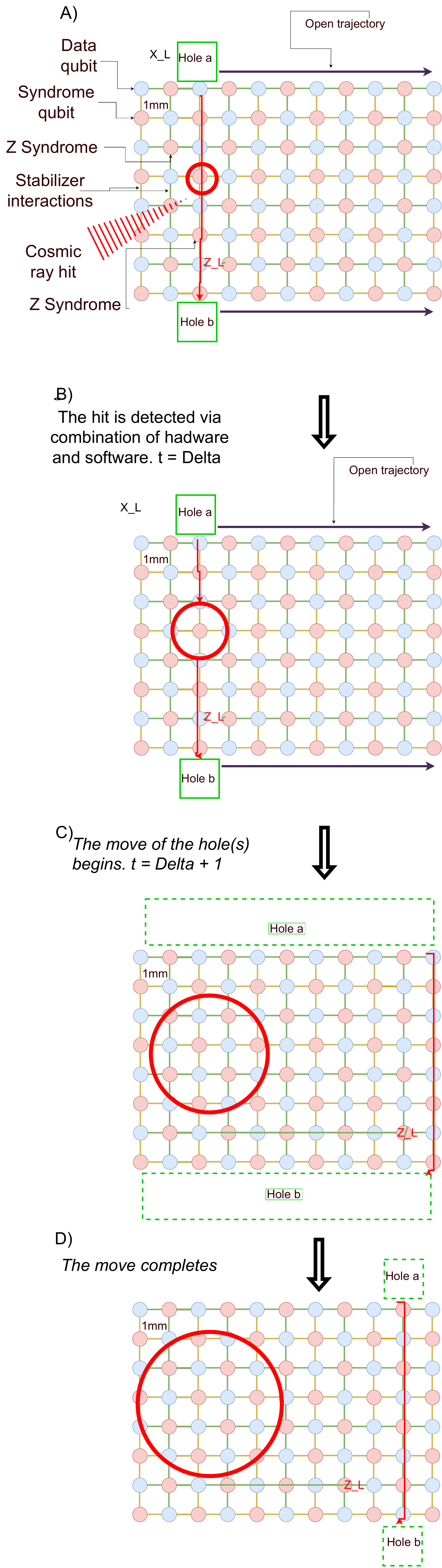}}
    \caption{With a two-hole logical qubit and a cosmic ray hit in between, both holes must move on clear, open rectilinear trajectories to flee the effects.}
    \label{fig:flight-whiteboard}
\end{figure}

If length $l$ is the physical qubit lattice spacing (e.g., marked as 1mm in Fig.~\ref{fig:flight-whiteboard}), increasing $l$ can slow the effective rate at which lattice cell-to-lattice cell propagation occurs, providing more time for software-based strategies to work and more time and distance for the phonons to dissipate. However, the desired physical interaction determines the preferred value of $l$. If time $t_c$ is the lattice cycle time, a shorter $t_c$ allows logical qubits to be moved more quickly, benefiting the software flight strategies below.

However, the time required for the circuit in Fig.~\ref{fig:measure-syndrome} is determined by the two-qubit gate times, which are determined by the relative qubit interaction frequency and strength and the measurement time, where longer measurement times may be higher fidelity.

Higher-fidelity physical operations reduce the required code distance, reducing the physical area of a logical qubit and hence the cross-section presented to potential incoming cosmic rays and shortening the hole move time. However, this reduction in code distance naturally reduces the physical distance between holes, and this tradeoff is investigated below.

\begin{figure}[t]
    \centering
    \includegraphics[width=8cm]{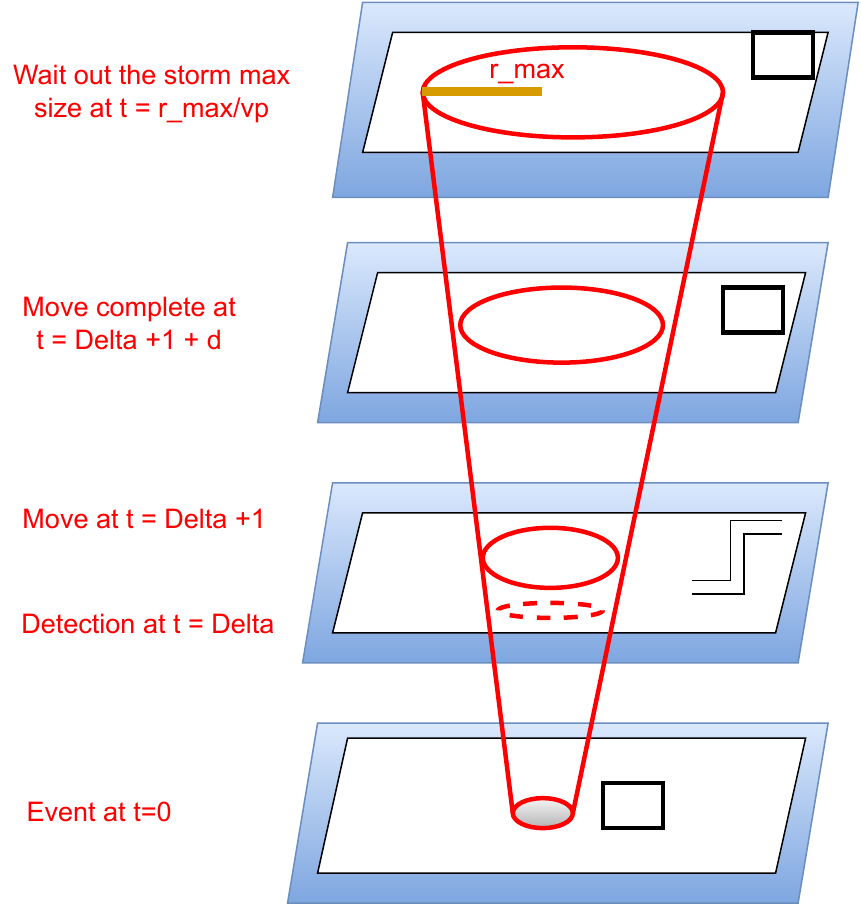}
    \caption{The hole can flee any distance along an available rectilinear path in a fixed amount of time. Max size at  $t = r_{\textrm{max}}/v_p$, move begins at $t=\Delta +1$, detection at $t=\Delta$, event at $t=0$.
}
   \label{fig:flight-cone-whiteboard-vf}
\end{figure}
\subsection{Software strategies}

Once the hardware requirements are met, the next step is software. As mentioned earlier, the idea is to move the logical qubit far enough to prevent the CRE from destroying it. In this paper, as in our previous work~\cite{PhysRevLett.129.240502}, we only consider errors caused by cosmic rays. We assume that the code distance has been set to suppress logical state errors due to ordinary decoherence thoroughly enough that we can ignore them.

We must be able to detect a cosmic ray strike as quickly as possible. While the details of such a determination remain to be determined, we can describe the general outline. As noted earlier, CREs preferentially result in $\ket{1}\rightarrow\ket{0}$ decay. If all of the qubits in a Z stabilizer decohere simultaneously, the syndrome qubit will still be measured as 0, indicating no error. However, the X stabilizer will result in 50/50 measurements of 0 and 1. This sudden asymmetry in syndrome values is a marker of CREs, especially if seen growing in a ring. In Fig.~\ref{fig:flight-cone-whiteboard-vf} and other places, we will designate the time at which a CRE is unequivocally detected either $\Delta$ or Delta.

Complementing the hardware strategies above, software (including the flexible elements of error correction) can contribute to solving the problem in ways we study in several sections. The simplest solution is to use a code distance that results in physical hole separation exceeding the dissipation radius $r_{\textrm{max}}$. However, the hole must also have a radius $>r_{\textrm{max}}$, or the phonon front can consume a single hole. Such a large physical radius is likely impractical, so we investigate a more dynamic approach in the next section.

\section{Flight: Detection and Escape}
\label{sec:fly}

Our proposed strategy is to \emph{flee the area}: move logical qubits far enough from the strike's epicenter to preserve our logical information. This is more easily accomplished using the 2-D surface code than other codes. In addition, as the state remains vulnerable during the move or while we maneuver other holes to allow us to move a particular hole, we must examine how well we can protect the logical qubit state in these cases.

Implementing this strategy will involve several phases. 
With time measured in units of the surface code cycle time $t_c$, assume detection of a hit in time $t= \Delta$ and an immediate and accurate response after that time. With phonon propagation velocity $v_p$, the phonon
front has advanced to a ring of radius $r_\Delta = \Delta v_p$ before software mitigation begins. The steps above are illustrated in Figs.~\ref{fig:flight-whiteboard} and  \ref{fig:flight-cone-whiteboard-vf}.

First, consider the unconstrained movement of a single hole. The complete sequence of events is as follows:
\begin{figure}[b]
    \centering
    \includegraphics[scale=0.36]{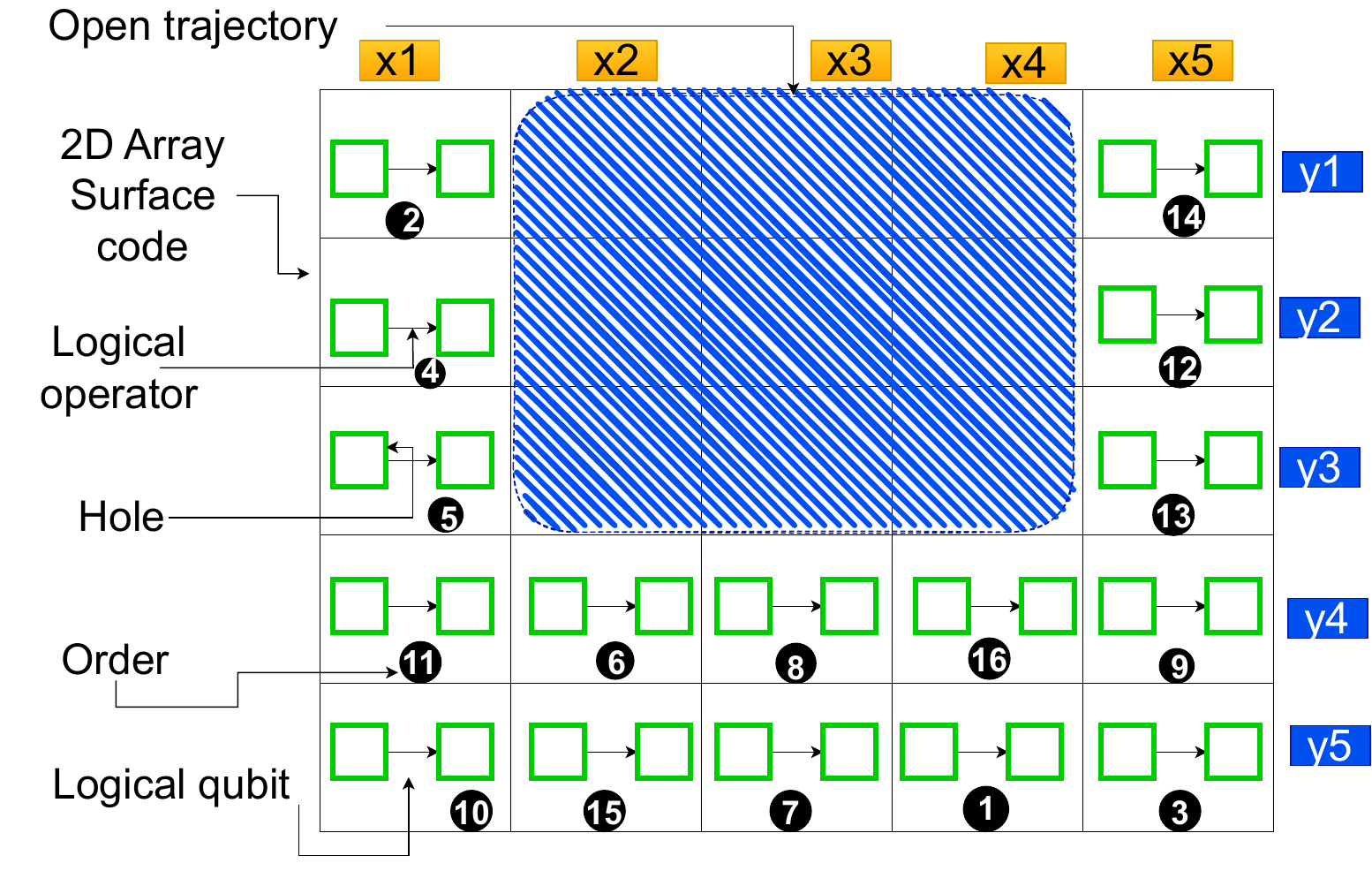}
    \caption{An example of multi-qubit mapping using 2-D array surface codes. In this mapping, the detection time determines the qubits' movement order. Wherever the CRE is detected, it enables easy qubit moves and limits time steps for moving our logical qubits.}
    \label{fig:example-multi}
\end{figure}
\begin{enumerate}
    \item $t=0$: The cosmic ray event occurs.
    \item $t=\Delta$: The strike is detected via a combination of hardware and software.
    \item $t=\Delta+1$: The move of the hole(s) begins.
    \item $t=\Delta+1+d$: The move completes.
    \item $t=\frac{r_{\textrm{max}}}{v_p}$: The logical data (hole) waits out the storm as the phonons dissipate. With appropriate management of resources, continuing the planned computation during the storm should be possible.
     \begin{figure*}[!h]
   \begin{center}
       \subfigure[Example of logical qubits moving after a cosmic ray event. Due to the first cosmic ray hit at position ($x_4$, $y_5$), the logical qubits at the positions ($x_3$, $y_4$), ($x_4$, $y_4$), and ($x_5$, $y_4$) must be moved.]{\label{fig:multi-logical-with-move}
       \includegraphics[scale=0.28]{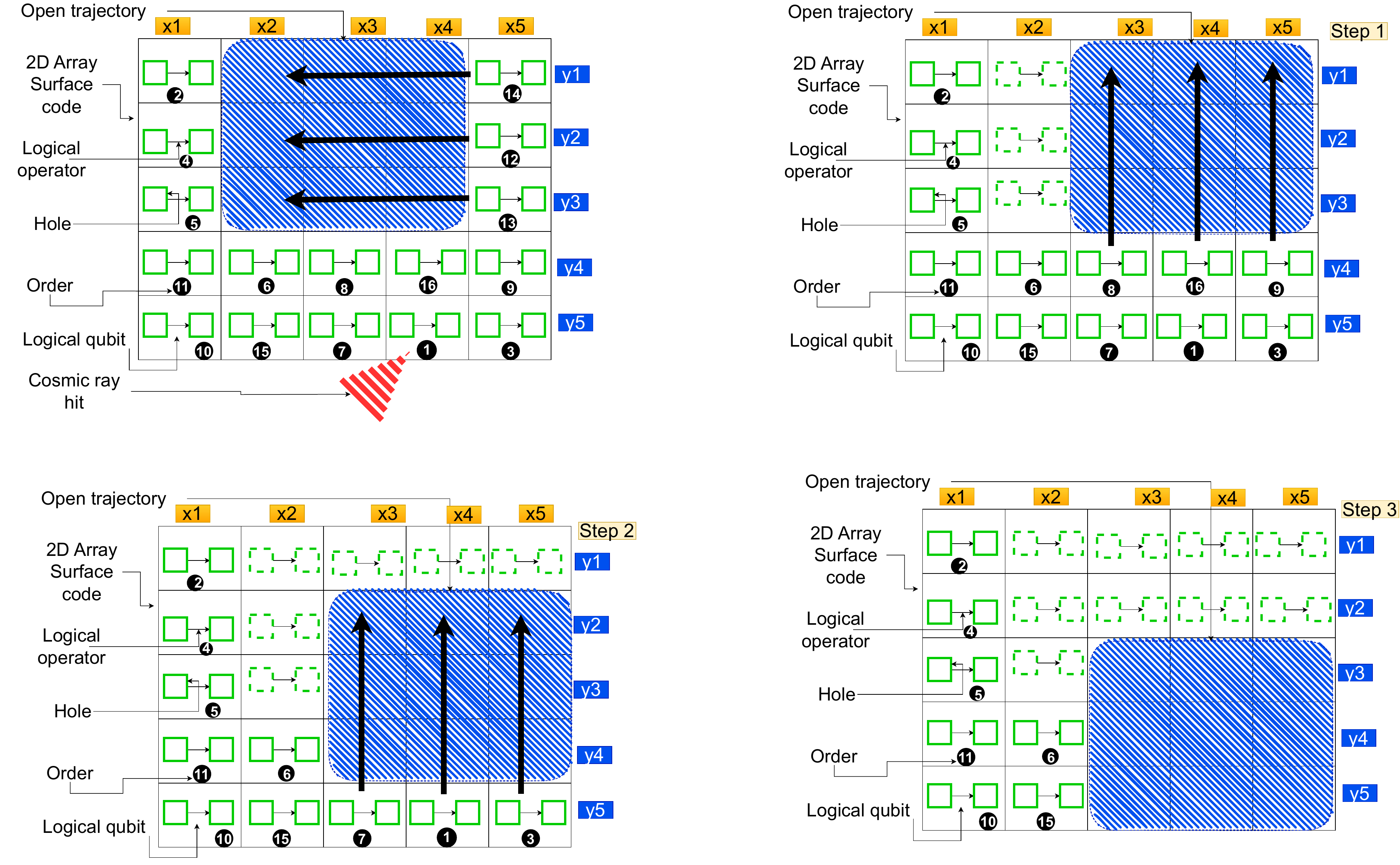}}
           
   \end{center}

    \begin{center}
    
     \subfigure[An example of a 2-D array surface code used to map many qubits. It is a repetition (top and bottom) and concatenation (left and right) of the simple mapping defined previously, and it can be extended continuously.]{\label{fig:example-multi-gen}
       \includegraphics[scale=0.4]{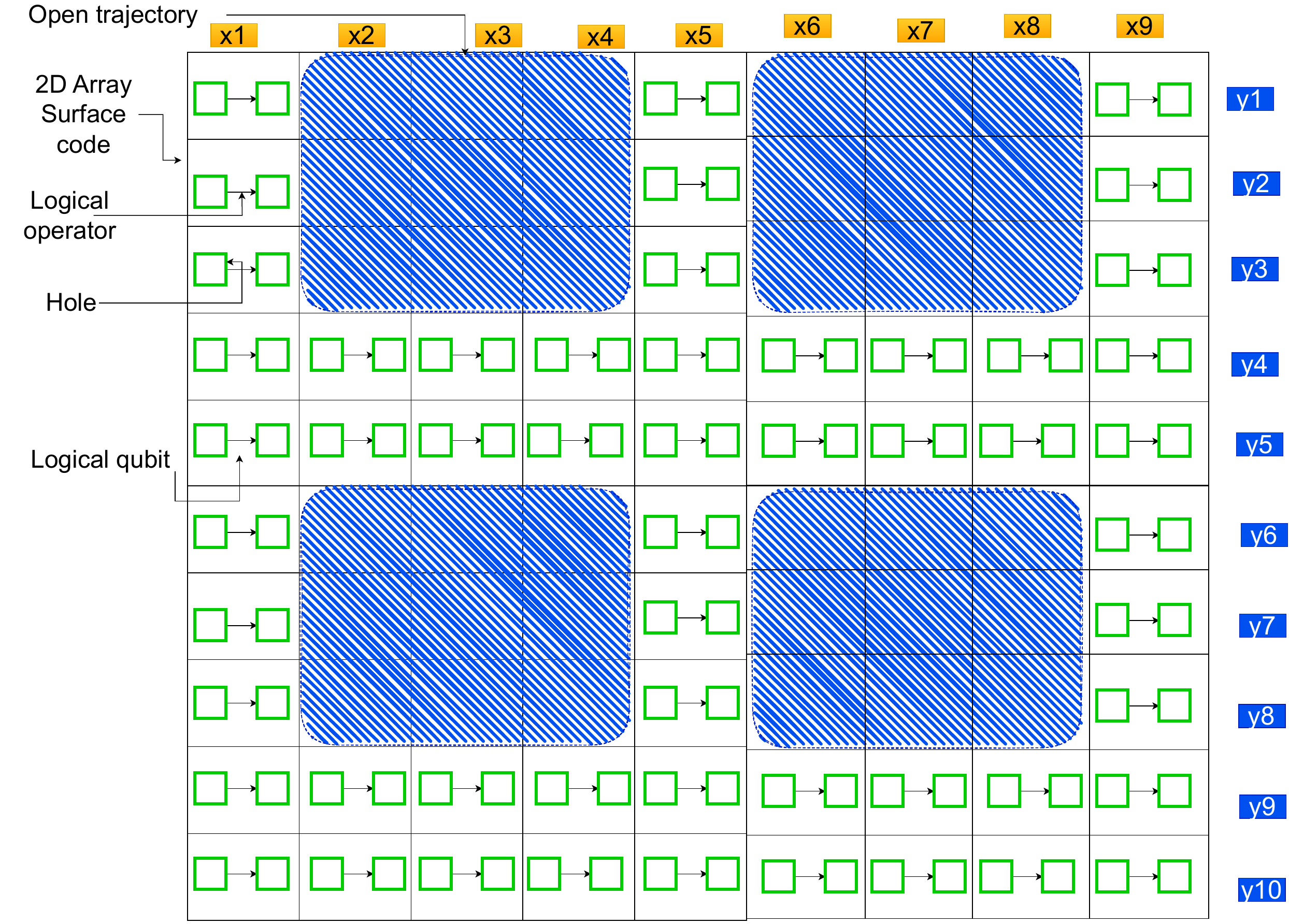}}

   \end{center}
    \caption{Examples of logical qubits moving and mapping a large number of qubits}
   \label{fig:wSBAlgo}
\end{figure*}
\clearpage
    \item The hole is returned to its original location, or other preparations are made for continuing the computation and weathering the next strike.
\end{enumerate}

During the movement of the hole (from $t=\Delta$ to $t=\Delta+1+d$), the phonon radius continues to grow, and the state is vulnerable until the move completes and the hole is far enough away for the phonons to dissipate.

Our approach must meet several constraints for this to work reliably:

\begin{enumerate}
    \item Phonon propagation must not overwhelm an entire logical qubit in less than $t=\Delta+1+d$.
    \item There must be somewhere for the logical qubit to move a sufficient distance away.
\end{enumerate}
The first constraint dictates the size of a logical qubit which depends on the speed of detection and movement. The lower the displacement speed, the greater the required length of the logical qubit. The second constraint dictates the radius $r_{\textrm{max}}$ that must be achieved in hardware and the density and placement of other logical qubits (discussed more in Sec.~\ref{sec:multiqubit}) used in the software. Hence we make the following assumptions:

\subsubsection{Assumptions} We assume the following conditions:
\label{subsec: assumptions}

\begin{subequations}

\begin{subequations}
    \begin{equation}
    v_p(\Delta + 1) < (x_0 - v_p(\Delta + 1)) + l(d - 1)
        \label{eq:condition-1}
    \end{equation}

    \begin{equation}
    \begin{split}
        r_{\max} < (x_0 - v_p(\Delta + 1))  + dl + l(d - 1)
        \label{eq:condition-2}
        \end{split}
    \end{equation}
\end{subequations}
where $r_{\max}$ is the maximum radius of the phonon ring, $dl$ is the distance traveled by the logical qubit between $t = \Delta + 1$ and $t=\Delta+1+d$ (the time the move is completed), $x_0$ is the distance from the cosmic ray location to the hole position at $t= 0$, $(x_0 - v_p(\Delta + 1))$ is the distance between the hole and the cosmic ray position when we start the move ($t = \Delta + 1$), $l$ is the physical qubit lattice spacing, and $d$ is the code distance (distance between the holes).  
\end{subequations}

Before the displacement process is triggered, the first condition \eqref{eq:condition-1} ensures that the logical qubit is not entirely compromised. Whenever the ring of phonons reaches its maximum radius, the condition \eqref{eq:condition-2} guarantee that the holes are at a sufficient distance or the ring phonons have consumed less than $d-1$ qubits.

Hence, (\eqref{eq:condition-1} \& \eqref{eq:condition-2}) should be satisfied for our mitigation technique to work.

\section{Multiple Logical Qubits}
\label{sec:multiqubit}
As mentioned in the previous section, in a multi-qubit environment, our ability to move a hole quickly depends on the position of the hole relative to the nearest open trajectory and the density of qubits we have mapped onto the surface. Hence, we propose an approach based on the mapping. This approach uses 2-D array surface codes in mapping against cosmic ray impacts.

If the logical qubits are closely packed, they can only be moved one at a time, increasing the time for the qubit nearest the strike site to flee by the number of qubits between it and the most immediate open space. In essence, we map the qubits so that whenever a CRE occurs, we can minimize the time steps affecting potentially vulnerable qubits' moving.

Figure \ref{fig:example-multi} illustrates a basic surface code mapping for several logical qubits. There is a displacement priority for each logical qubit based on the position and detection order of the CRE. depending on this mapping, no matter where the CRE is detected, the number of time steps in our surface code array does not exceed $3$ as shown in Fig.~ \ref{fig:multi-logical-with-move}. Using this simple mapping as a starting point, Figure \ref{fig:example-multi-gen}  proposes a generalization. The latter is nothing more than a concatenation (left and right) and layering (top and bottom) of Fig.~\ref{fig:example-multi}. Moreover, it can be extended continuously.


       



   

\begin{figure*}[!h]

   \begin{center}
       
    \includegraphics[scale=0.6]{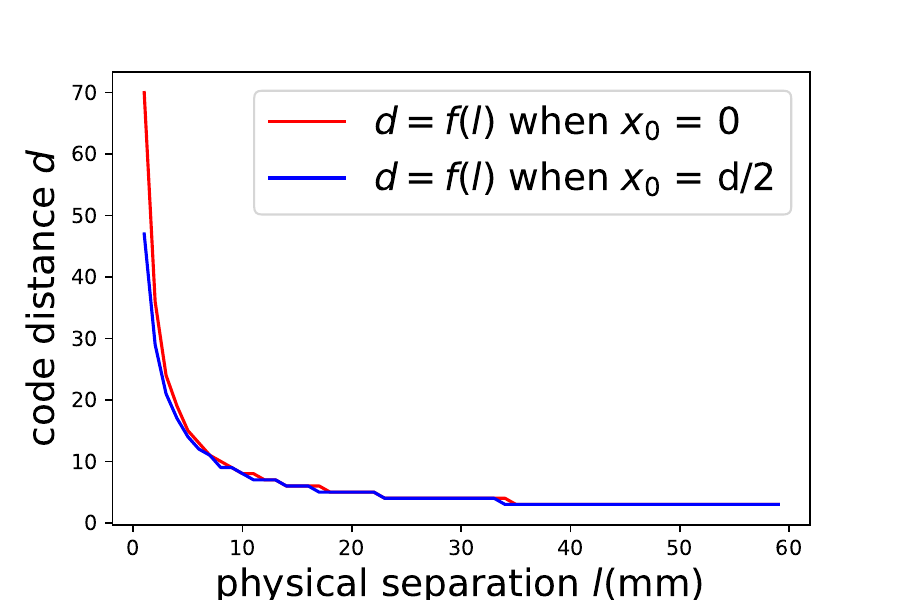}


   \end{center}
   \caption{Variation of the necessary \textit{code distance d} when the cosmic ray events hit exactly halfway between the holes ($x_0=d/2$) and close to one hole ($x_0=0$) depending on the physical distance ($l$) between qubits where $r_{\max} =  63$mm, $\Delta = 1\mu$sec, 
    $dl = 1$mm.}
   \label{fig:set-d(l)}

   
\end{figure*}

\begin{figure*}[!h]

   \begin{center}
    \subfigure[Variation of the \textit{distance d} when the CRE hits exactly halfway between the holes depending on the maximum radius of the  phonon ring $r_{\max}$ where ($l = 1$mm) ]{\label{fig:m-l1-d(rmax)}
    \includegraphics[scale=0.38]{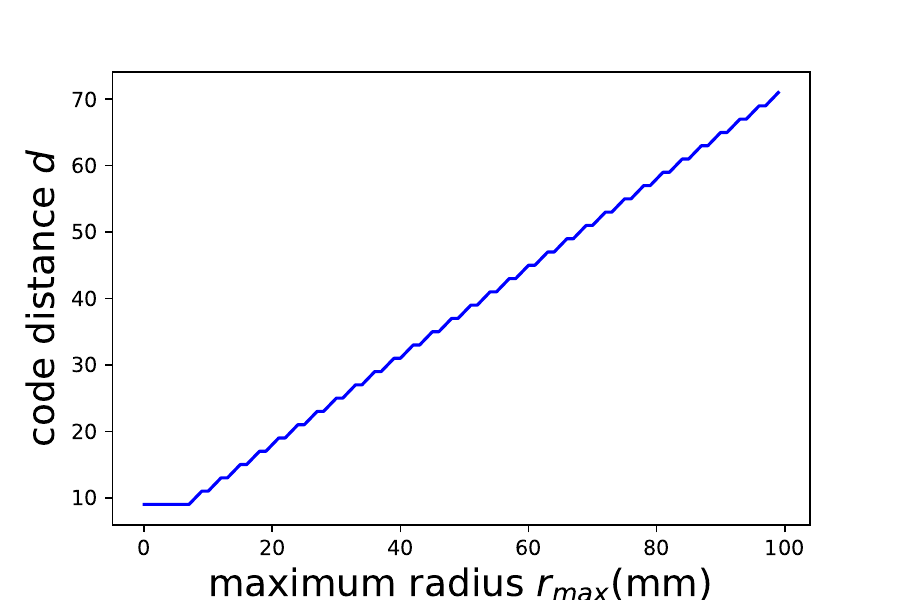}}
    \hfill
     \subfigure[Variation of the \textit{distance d} when the CRE hits exactly halfway between the holes depending on the maximum radius of the  phonon ring $r_{\max}$ where ($l = 5$mm)]{\label{fig:m-l5-d(rmax)}
       \includegraphics[scale=0.38]{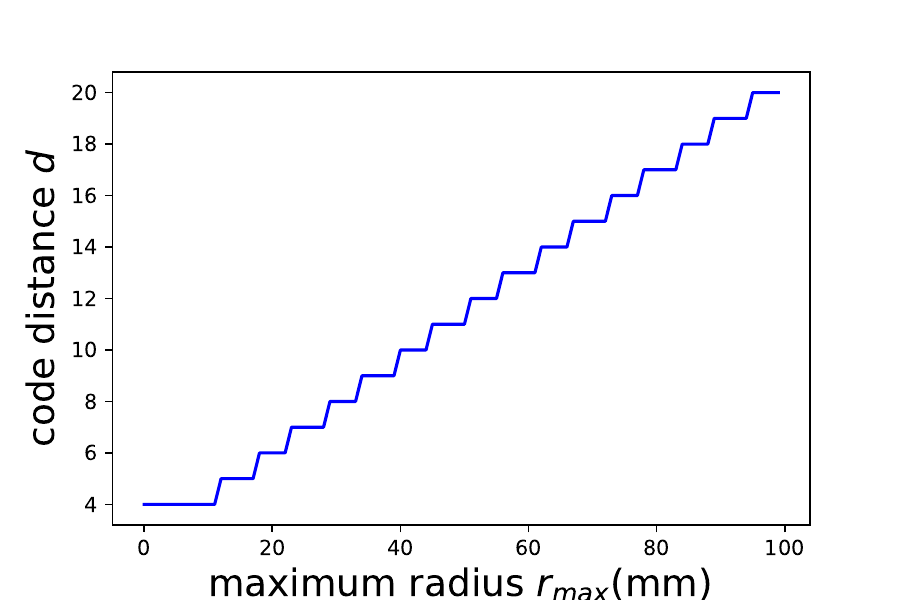}}
       \hfill
     \subfigure[Variation of the \textit{distance d} when the CRE hits exactly halfway between the holes depending on the maximum radius of the  phonon ring $r_{\max}$ where ($l = 10$mm)]{\label{fig:m-l10-d(rmax)}
       \includegraphics[scale=0.38]{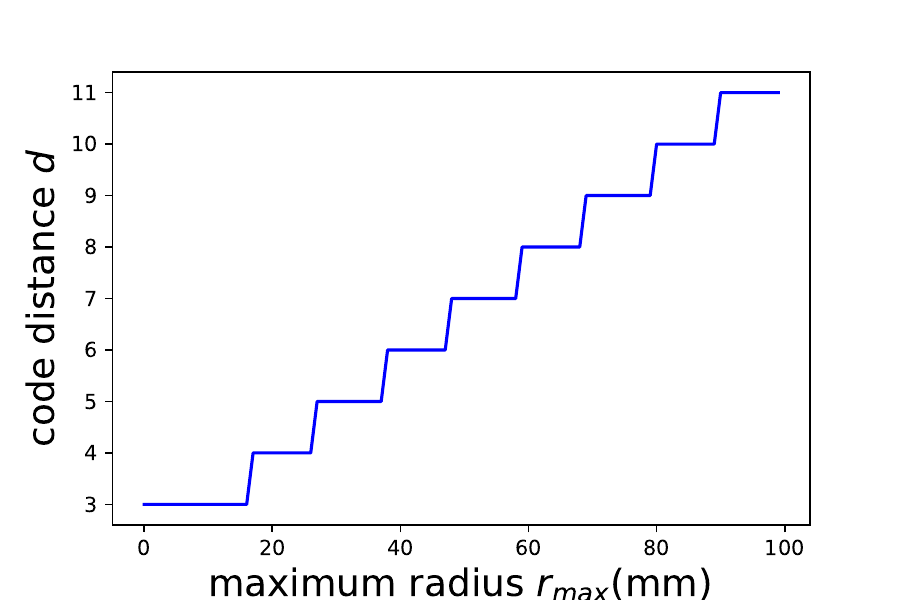}}
       
       \subfigure[Variation of the \textit{distance d} when the CRE hits close to one hole depending on the maximum radius of the phonon ring $r_{\max}$ where ($l = 1$mm) ]{\label{fig:n-l1-d(rmax)}
    \includegraphics[scale=0.38]{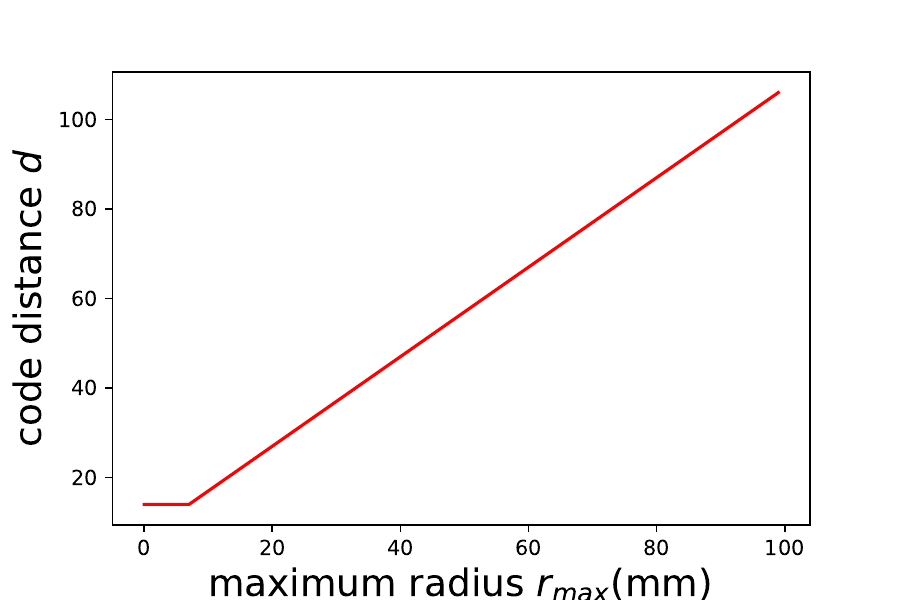}}
    \hfill
     \subfigure[Variation of the \textit{distance d} when the CRE hits close to one hole depending on the maximum radius of the phonon ring $r_{\max}$ where ($l = 5$mm)]{\label{fig:n-l5-d(rmax)}
       \includegraphics[scale=0.38]{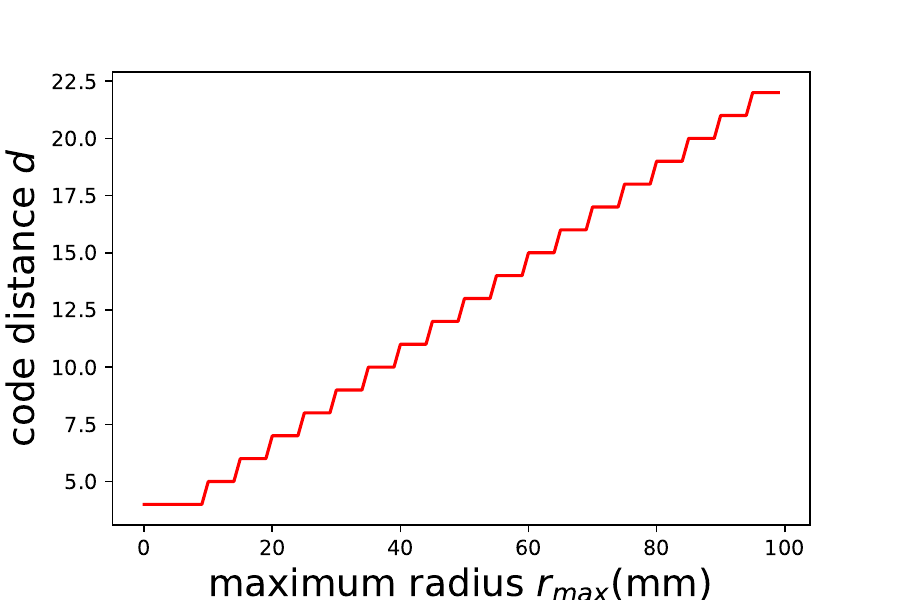}}
       \hfill
     \subfigure[Variation of the \textit{distance d} when the CRE hits close to one hole depending on the maximum radius of the phonon ring $r_{\max}$ where ($l = 10$mm)]{\label{fig:n-l10-d(rmax)}
       \includegraphics[scale=0.38]{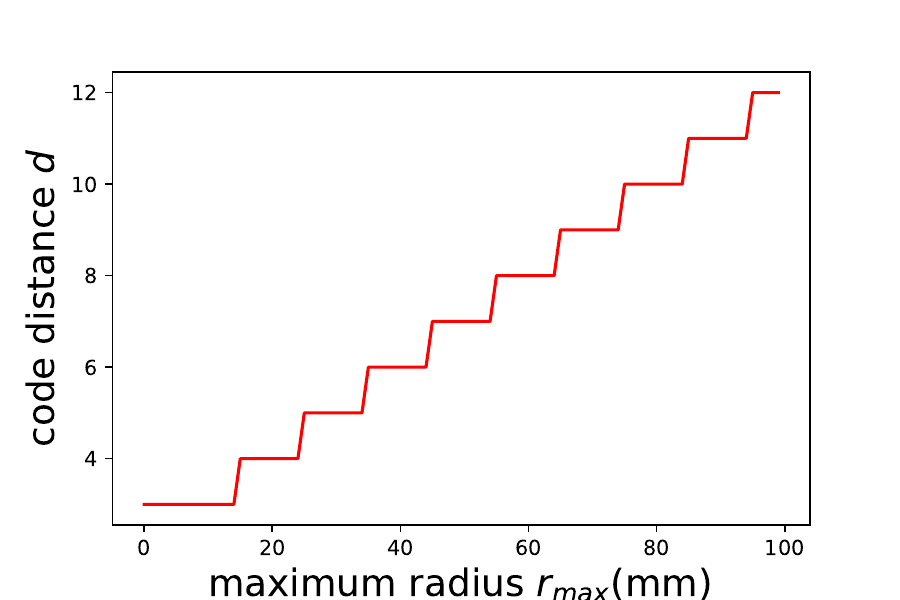}}


   \end{center}

   \begin{center}
    \subfigure[Variation of the \textit{distance d} when the CRE hits exactly halfway between the holes and close to one hole depending on the maximum radius of the phonon ring $r_{\max}$ where ($l = 1$mm) ]{\label{fig:combine-l1-d(rmax)}
    \includegraphics[scale=0.38]{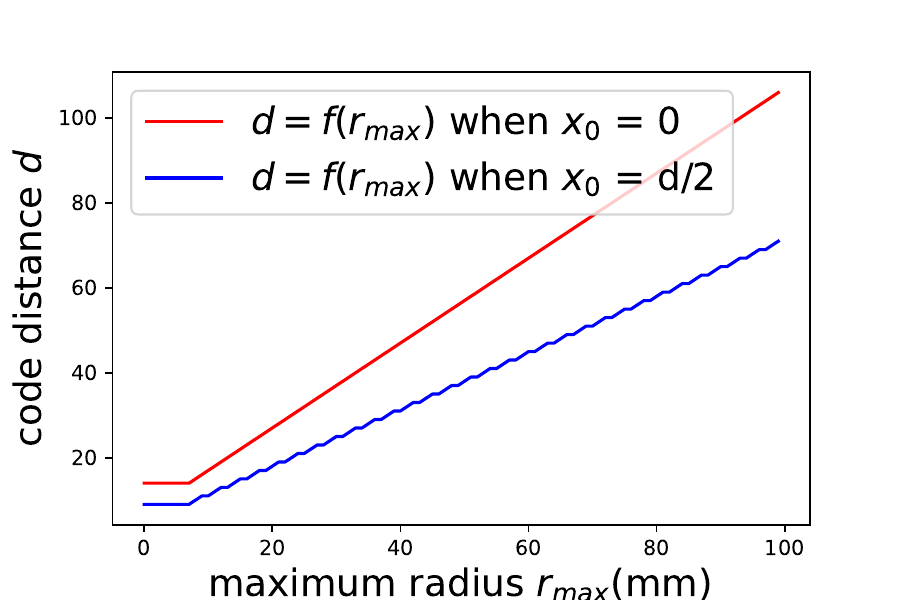}}
    \hfill
     \subfigure[Variation of the \textit{distance d} when the CRE hits exactly halfway between the holes and close to one hole depending on the maximum radius of the phonon ring $r_{\max}$ where ($l = 5$mm)]{\label{fig:combine-l5-d(rmax)}
       \includegraphics[scale=0.38]{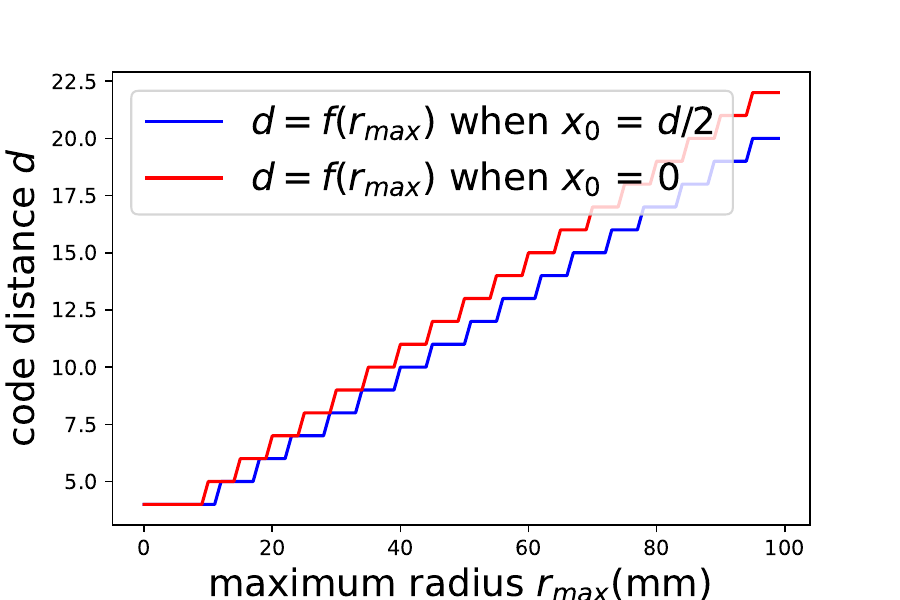}}
       \hfill
     \subfigure[Variation of the \textit{distance d} when the CRE hits exactly halfway between the holes and close to one hole depending on the maximum radius of the phonon ring $r_{\max}$ where ($l = 10$mm)]{\label{fig:combine-l10-d(rmax)}
       \includegraphics[scale=0.38]{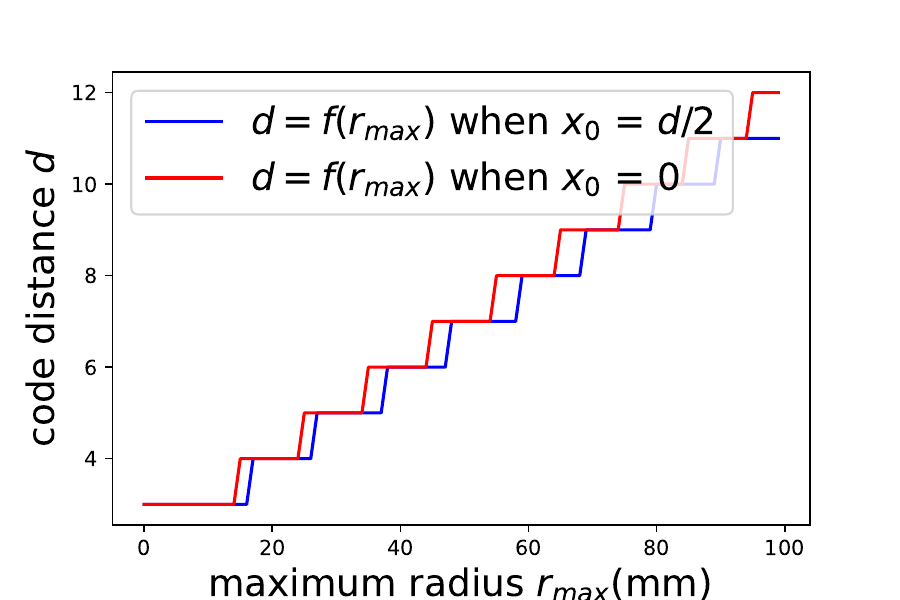}}


   \end{center}
   \caption{Variation of the \textit{distance d} when the cosmic ray events hit exactly halfway between the holes and close to one hole depending on the maximum radius of the phonon ring $r_{\max}$ where $\Delta = 1\mu$sec, $r_{\max} = [1, 100]$,  and $dl = 1$mm}
   \label{fig:set-d(r_max)}

   
\end{figure*}

\section{Evalauation}
\label{sec:eval}
On the one hand, the main objective of this analysis is to show that it is possible to satisfy the necessary conditions (\eqref{eq:condition-1} \& \eqref{eq:condition-2}) of our technique for mitigation of CREs. On the other hand, we illustrate how the distance $d$ of the code varies based on the space between qubits $l$, the maximum radius of the phonon $r_{\max}$, and the CRE detection time $\Delta$. We consider two basic CRE cases:

\begin{itemize}
    \item far from both holes (exactly halfway between): $x_0 = \frac{d}{2}$
    \item close to or overlapping with one hole: $x_0 = 0$
\end{itemize}
\subsection{Simulation settings}
We used a Python tool (CpModel) for the simulation that efficiently solves and evaluates problems with constraints.  We consider the following parameters:
\begin{itemize}
    \item When we evaluate $d$ depending on the value of $l$: \begin{equation}
    \begin{split}
        r_{\max} =  63\text{mm}, \Delta = [1, 25], 
    dl = [1, 1000000] \  and \\ l = [1, 60]
    \end{split}
    \label{eq:d(l)}
    \end{equation}
    \item When we evaluate $d$ depending on the value of $r_{\max}$: 
    \begin{equation}
    \begin{split}
        r_{\max} =  [1, 100], \Delta = [1, 25], 
    dl = [1, 1000000] \  and \\ l = 1\text{mm}, 5\text{mm}, and 10\text{mm}
    \end{split}
    \label{eq:d(rmax)}
    \end{equation}
    \item When we evaluate $d$ depending on the value of $\Delta$: 
    
    \begin{equation}
    \begin{split}
         r_{\max} = 63mm, \Delta = [1, 25],\\
    dl = [1, 1000000] \  and \ l = 1mm, 5mm, \ and \ 10mm
    \end{split}
    \label{eq:d(delta)}
    \end{equation}
\end{itemize}
We chose $\Delta$ and $dl$ randomly from the intervals defined in \eqref{eq:d(l)}, \eqref{eq:d(rmax)}, and \eqref{eq:d(delta)}.

The speed of the phonons propagating through the chip varies depending on the material and ranges between about $1-8$ km/s~\cite{martiniss415342021saving}. We have used the $2.5$mm$/\mu$sec in our simulations as in Silicon, the most popular material for substrates. We assume also that with a hardware strategy, the phonon propagation radius will be limited to $r_{\max} = 63$ mm. In addition, we have selected $l = 1$mm, as in the Google Sycamore quantum processor~\cite{mcewen2021resolving}.

\subsection{Evaluation and Discussions}
\begin{figure*}[!h]

   \begin{center}
    \subfigure[Variation of the \textit{distance d} when the CRE hits exactly halfway in between the holes depending on the detection time $\Delta$ where ($l = 1$mm) ]{\label{fig:m-l1-d(delta)}
    \includegraphics[scale=0.38]{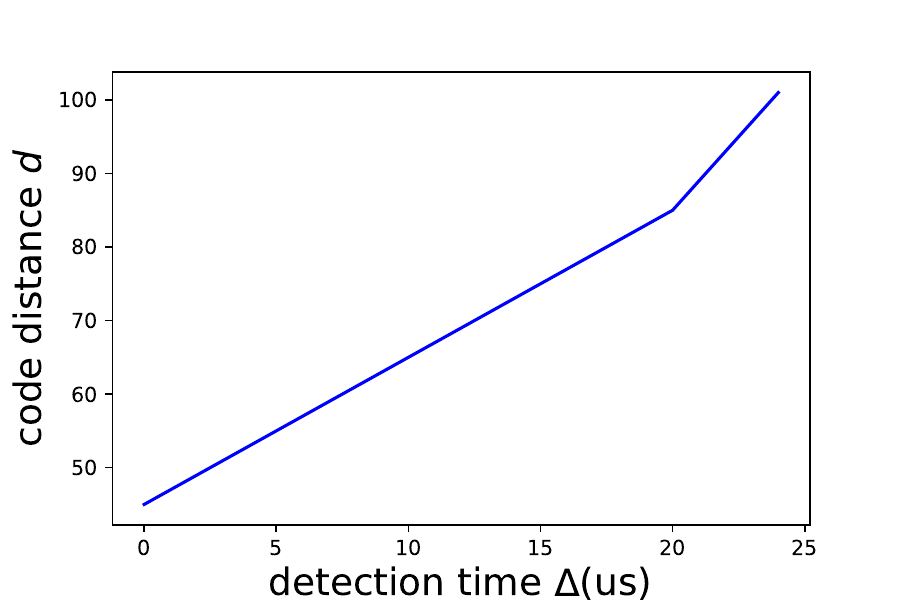}}
    \hfill
     \subfigure[Variation of the \textit{distance d} when the CRE hits exactly halfway in between the holes depending on the detection time $\Delta$ where ($l = 5$mm)]{\label{fig:m-l5-d(delta)}
       \includegraphics[scale=0.38]{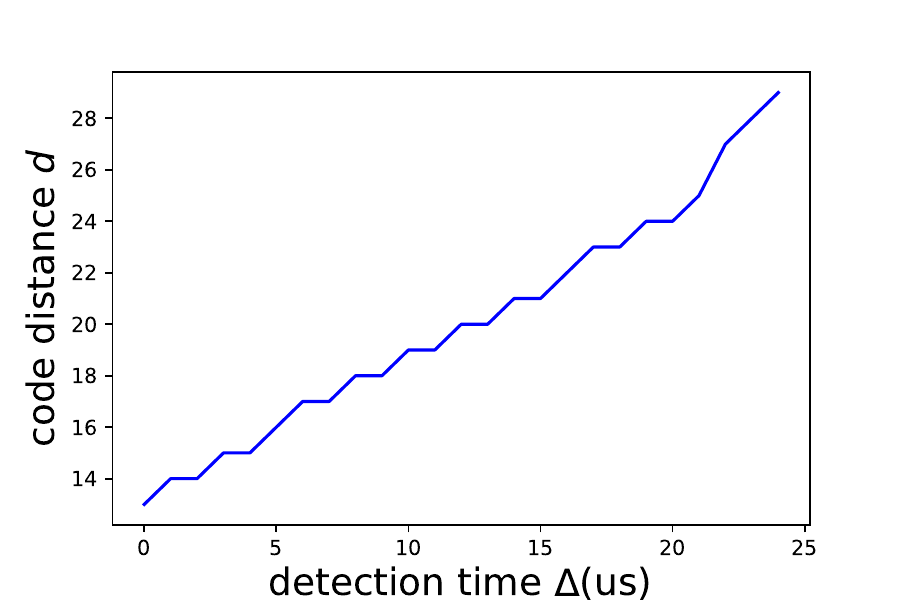}}
       \hfill
     \subfigure[Variation of the \textit{distance d} when the CRE hits exactly halfway in between the holes depending on the detection time $\Delta$ where ($l = 10$mm)]{\label{fig:m-l10-d(delta)}
       \includegraphics[scale=0.38]{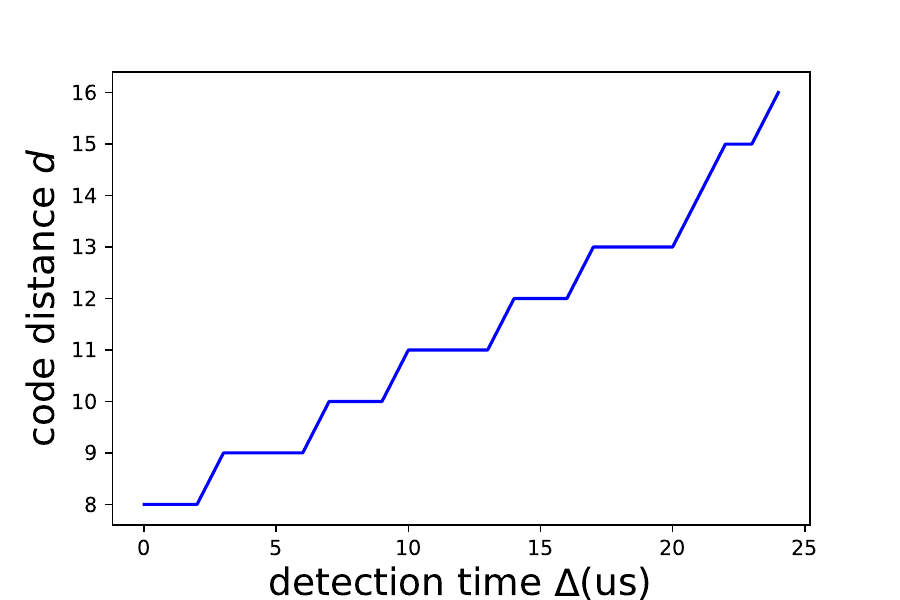}}
       
       \subfigure[Variation of the \textit{distance d} when the CRE hits close to one hole depending on the detection time $\Delta$ where ($l = 1$mm) ]{\label{fig:n-l1-d(delta)}
    \includegraphics[scale=0.38]{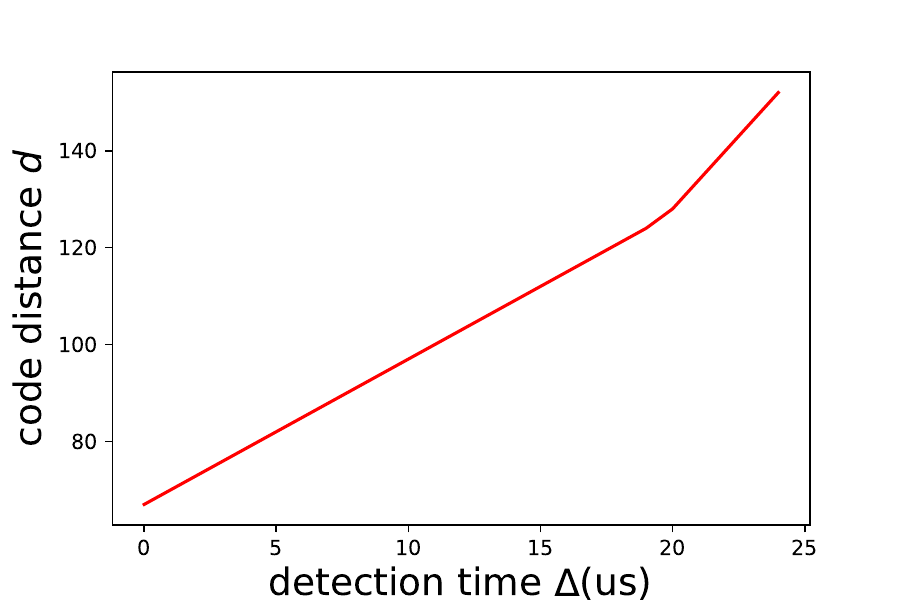}}
    \hfill
     \subfigure[Variation of the \textit{distance d} when the CRE hits close to one hole depending on the detection time $\Delta$ where ($l = 5$mm)]{\label{fig:n-l5-d(delta)}
       \includegraphics[scale=0.38]{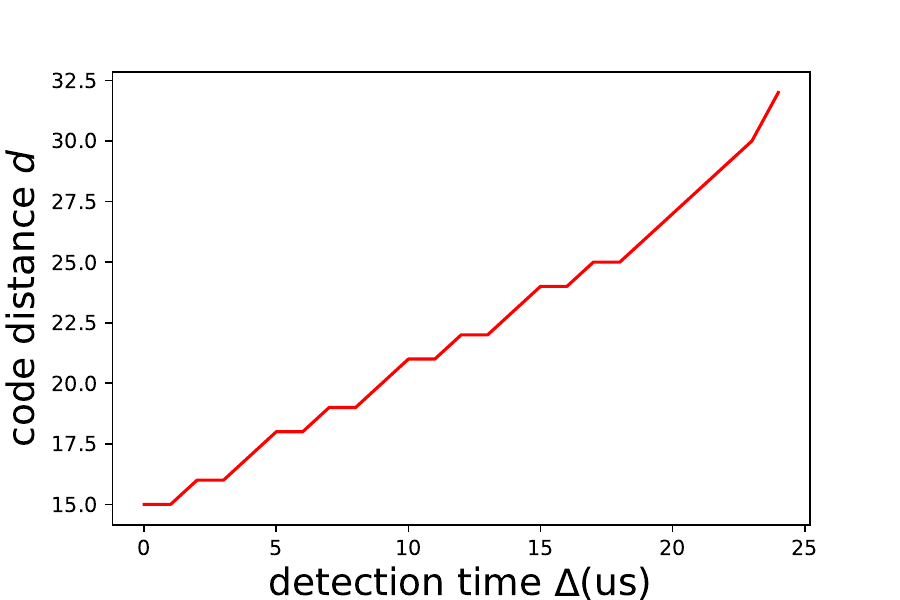}}
       \hfill
     \subfigure[Variation of the \textit{distance d} when the CRE hits close to one hole depending on the detection time $\Delta$ where ($l = 10$mm)]{\label{fig:n-l10-d(delta)}
       \includegraphics[scale=0.38]{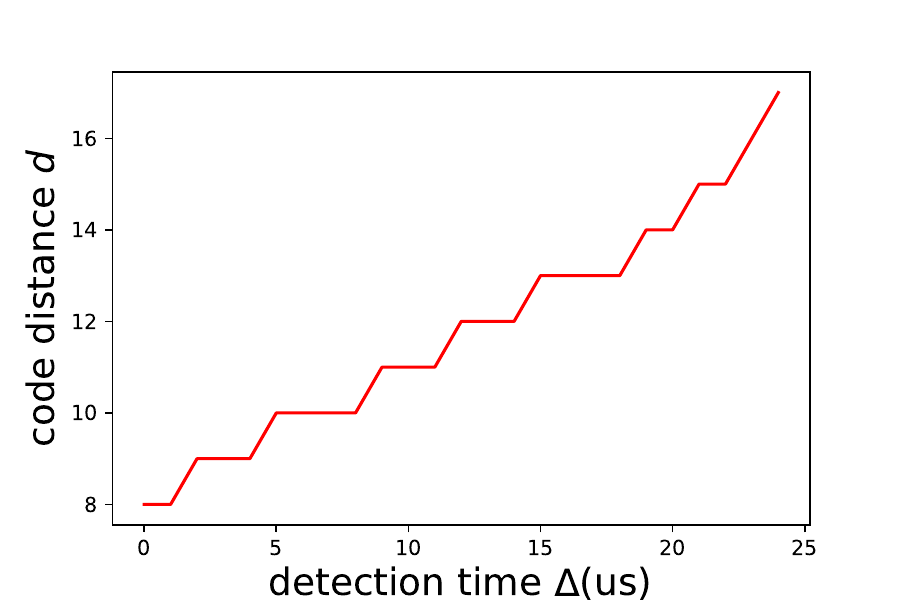}}


   \end{center}

   \begin{center}
    \subfigure[Variation of the \textit{distance d} when the CRE hits exactly halfway in between the holes and close to one hole depending on the detection time $\Delta$ where ($l = 1$mm) ]{\label{fig:combine-l1-d(delta)}
    \includegraphics[scale=0.38]{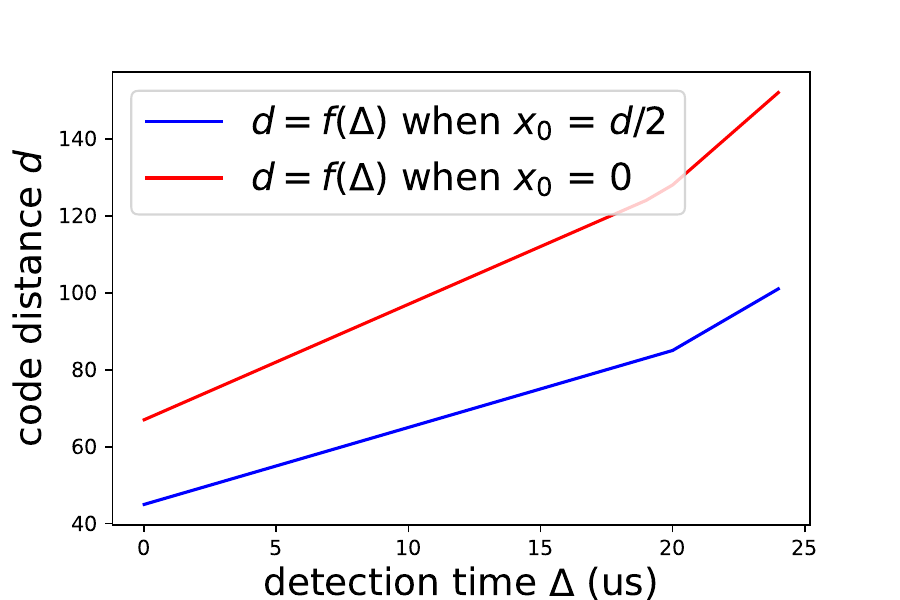}}
    \hfill
     \subfigure[Variation of the \textit{distance d} when the CRE hits exactly halfway in between the holes and close to one hole depending on the detection time $\Delta$ where ($l = 5$mm)]{\label{fig:combine-l5-d(delta)}
       \includegraphics[scale=0.38]{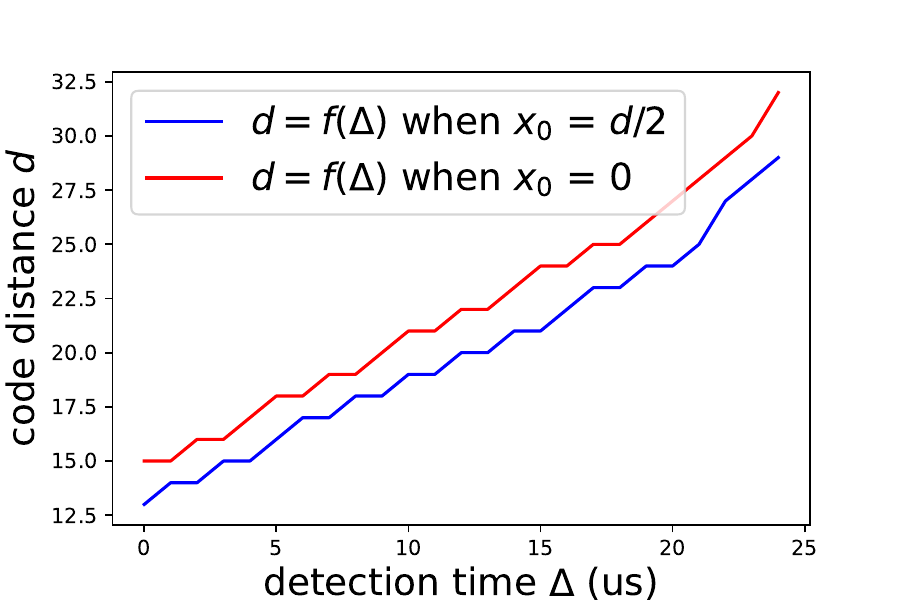}}
       \hfill
     \subfigure[Variation of the \textit{distance d} when the CRE hits exactly halfway in between the holes and close to one hole depending on the detection time $\Delta$ where ($l = 10$mm)]{\label{fig:combine-l10-d(delta)}
       \includegraphics[scale=0.38]{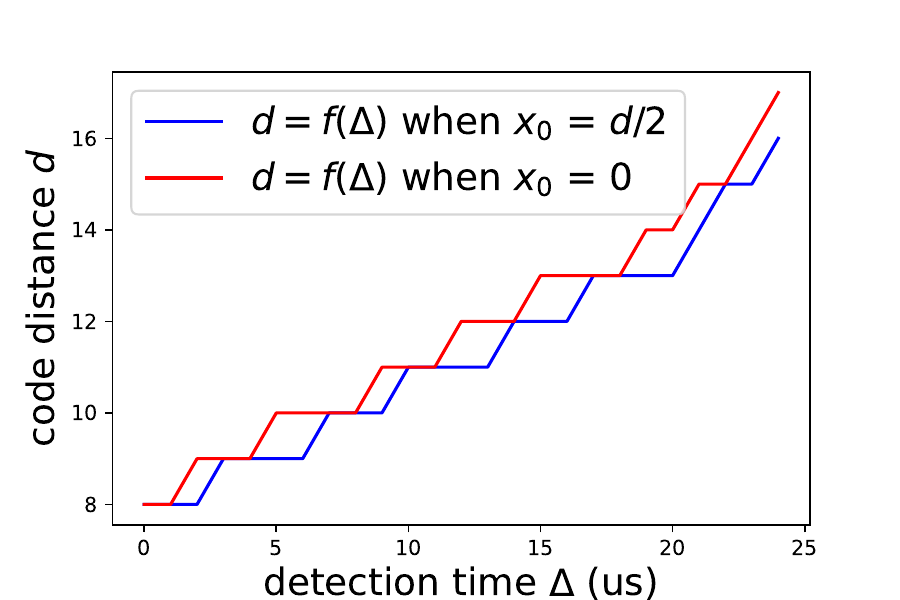}}


   \end{center}
   \caption{Variation of the \textit{distance d} when the cosmic ray events hit exactly halfway in between the holes and close to one hole depending on the detection time $\Delta$ where  $r_{\max} = 63$mm,  and $dl = 1$mm}
   \label{fig:set-d(delta)}

   
\end{figure*}

Based on the assumptions (\eqref{eq:condition-1} \& \eqref{eq:condition-2}), after determining how long it takes to detect a CRE and flee from it ($\Delta + 1$), we choose the code distance $d$ such that we can always get away from the CRE and never lose a logical qubit. Figures \ref{fig:set-d(l)}, \ref{fig:set-d(r_max)}, and \ref{fig:set-d(delta)} illustrate the possible choices of the distances $d$ under several scenarios.

As you can observe in Fig.~\ref{fig:set-d(l)}, the minimum necessary distance $d$ of the code varies inversely to the separation between the qubits $l$. When the inter-qubit spacing is small, the required code distance $d$ becomes large. This value decreases considerably when the spacing increases. This corresponds to our expectations. Indeed, when there is enough space between the qubits, the probability of compromise of the data becomes low, allowing the choice of a small code distance $d$.

Figure~\ref{fig:set-d(l)} compares the variation of the code distance when the cosmic ray hits in between the holes and near a hole. We observe that for a small spacing, the distance is wider when the cosmic ray strikes close to the hole than when it occurs in the middle. However, as the spacing increases, the distance tends to be the same in both cases. In other words, depending on the spacing between the qubits, we can set a distance that protects the logical qubit regardless of where the comic ray occurs. 

In Figs.~\ref{fig:m-l1-d(rmax)}, \ref{fig:m-l5-d(rmax)}, and \ref{fig:m-l10-d(rmax)}, we can see the variation of the code distance $d$ depending on the maximum radius of the phonon ring $r_{\max}$, where $l = 1 $mm, $l = 5 $mm, and $l = 10 $mm, respectively, when the cosmic ray hits halfway between the holes. Figs.~\ref{fig:n-l1-d(rmax)}, \ref{fig:n-l5-d(rmax)}, and \ref{fig:n-l10-d(rmax)} show the variation of the distance $d$ of the code when the cosmic ray strikes near one hole based on the maximum radius of the phonon ring $r_{\max}$, where $l = 1$mm, $l = 5$mm, and $l = 10$mm, respectively. 

As the maximum radius of the phonon ring increases, the required distance $d$ increases linearly as illustrated in Figs.~\ref{fig:m-l1-d(rmax)}, \ref{fig:m-l5-d(rmax)}, \ref{fig:m-l10-d(rmax)}, \ref{fig:n-l1-d(rmax)}, \ref{fig:n-l5-d(rmax)}, and \ref{fig:n-l10-d(rmax)}. In other words, the more the phonon propagation can be limited, the lower a code distance is required. However, we also observe that this distance decreases when we gradually increase the spacing between the qubits. In fact, as the spacing $l$ varies from $1$mm to $5$mm, the distance changes from $[10, 70]$ to $[4, 20]$. It also varies from $[10, 70]$ to $[3, 11]$ when the spacing is equal to $1$mm, $10$mm. So once we master the maximum radius of the phonon ring on the hardware side, we can define an appropriate distance by considering the spacing. A larger spacing also makes the same distance susceptible to protecting our logical qubit for various maximum phonon ring radiuses. This justifies the horizontal lines that widen more and more. Moreover, the case where the cosmic ray hits near one hole requires a vast distance compared to when it occurs halfway between the holes as illustrated in Figs.~\ref{fig:combine-l1-d(rmax)}, \ref{fig:combine-l5-d(rmax)}, and \ref{fig:combine-l10-d(rmax)}. When the spacing between qubits increases, this difference becomes less important, so the curves tend to overlap.

\subsection{Evaluation of the probability of failure}
Given our system's mapping, the logical qubit remains vulnerable when a CRE occurs inside one of these holes. Let's call this event $\mathcal{A}$. We denote by $\mathcal{E}$ the event of a CRE affecting $d-1$ qubits. Thus the evaluation of the probability of the event ($\mathcal{E} \wedge \mathcal{\overline{A}}$) allows us to estimate the probability of success of our mitigation technique. $(\mathcal{E} \wedge \mathcal{\overline{A}})$ means that the CRE occurs somewhere other than inside a logical qubit's hole and affects less than $d-1$ of its qubits. Since $\mathcal{E}$ and $\mathcal{\overline{A}}$ are two independent events then $\mathcal{P}(\mathcal{E} \wedge \mathcal{\overline{A}}) = \mathcal{P}(\mathcal{E}) \times \mathcal{P}(\overline{A})$ where $\mathcal{P}(\overline{A}) = 1 - \mathcal{P}(\mathcal{A})$.

\begin{figure}[!t]
    \centering
    \includegraphics[scale=0.2]{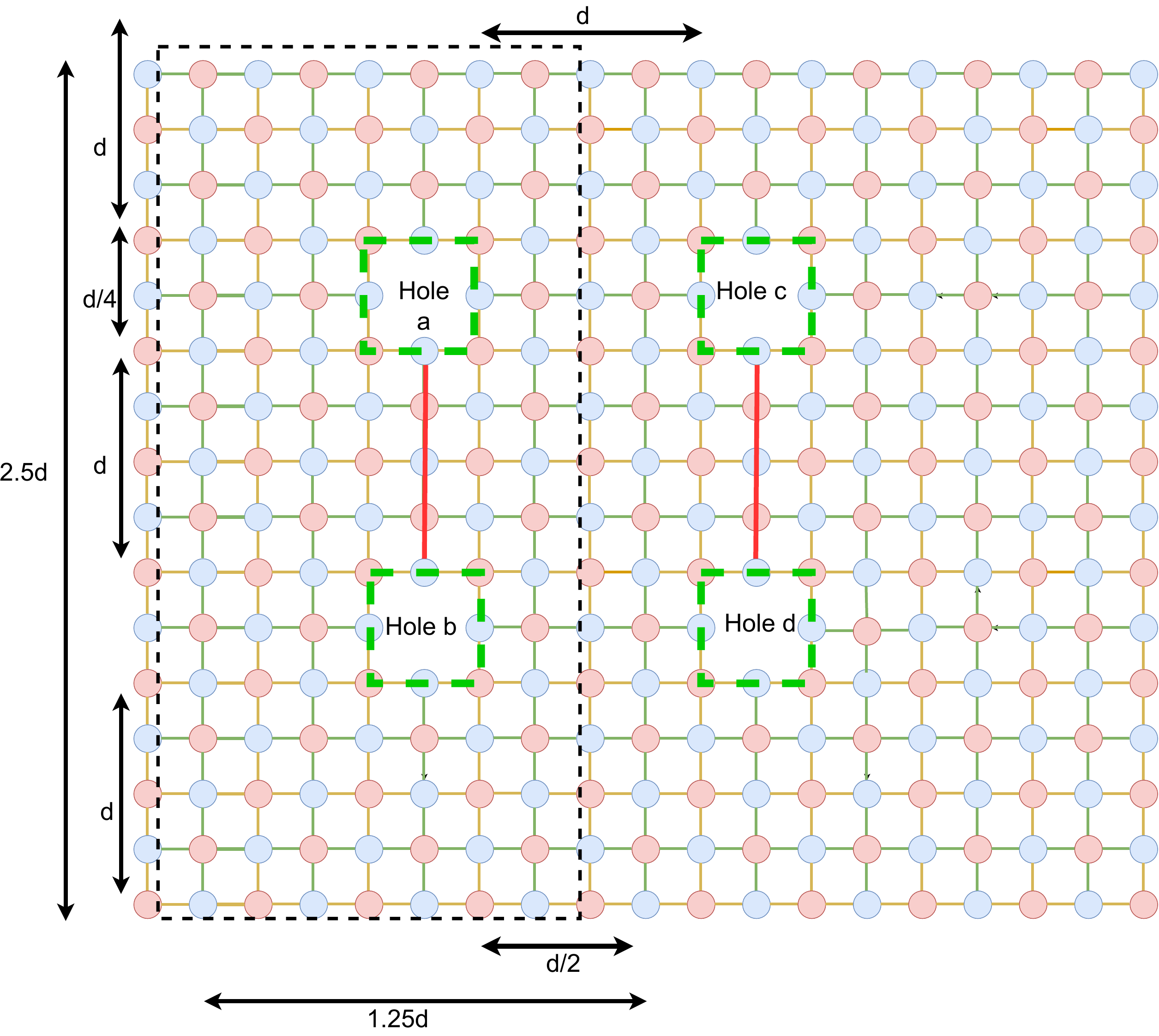}
    \caption{Evaluation of the probability of failure where the cosmic ray events hit close to or overlapping with one hole.}
    \label{fig:eval-design}
\end{figure}

Suppose, for instance, we have the mapping in Fig.~\ref{fig:eval-design}: where the logical qubits are separated by $d$, and the dashed line designates a region where a logical qubit might be accommodated. ($\frac{10d}{4} \times \frac{5d}{4}$)/($d/4$ $\times$ $d/4$) is the maximum number of holes (each square $d/4$ by $d/4$) we can have in this frame. Consequently, the probability of a CRE occurring inside a hole is equal to $\mathcal{P}(\mathcal{A}) = \frac{2}{50}$ (a logical qubit consists of two holes). Let us evaluate $P(\mathcal{E})$. As in our previous paper \cite{PhysRevLett.129.240502}, we consider that the cosmic ray event distribution follows the Poisson distribution. Therefore $\mathcal{N}(t) \xrightarrow{} \mathcal{P}(\lambda\tau)$ and \[\mathcal{P}(\mathcal{E}) = \mathcal{P}[\mathcal{N}(t) < d -1] = e^{-\lambda\tau} \sum^{d -2}_{k=0} \frac{(\lambda\tau)^k}{k!} \], where $\lambda$ represents the chip's CRE rate, and $\tau$ is the time needed to move our logical qubit to a safe location. Hence \[\mathcal{P}(\mathcal{E} \wedge \mathcal{\overline{A}}) =   (1 - \frac{2}{50})\times(e^{-\lambda\tau} \sum^{d -2}_{k=0} \frac{(\lambda\tau)^k}{k!})\] So, we can suppress the failure rate caused by a CRE in the chip to \[1 - \mathcal{P}(\mathcal{E} \wedge \mathcal{\overline{A}}) = 1 - (1 - \frac{2}{50})\times(e^{-\lambda\tau} \sum^{d -2}_{k=0} \frac{(\lambda\tau)^k}{k!})\]. 

\begin{figure}[!t]
    \centering
    \includegraphics[scale=0.6]{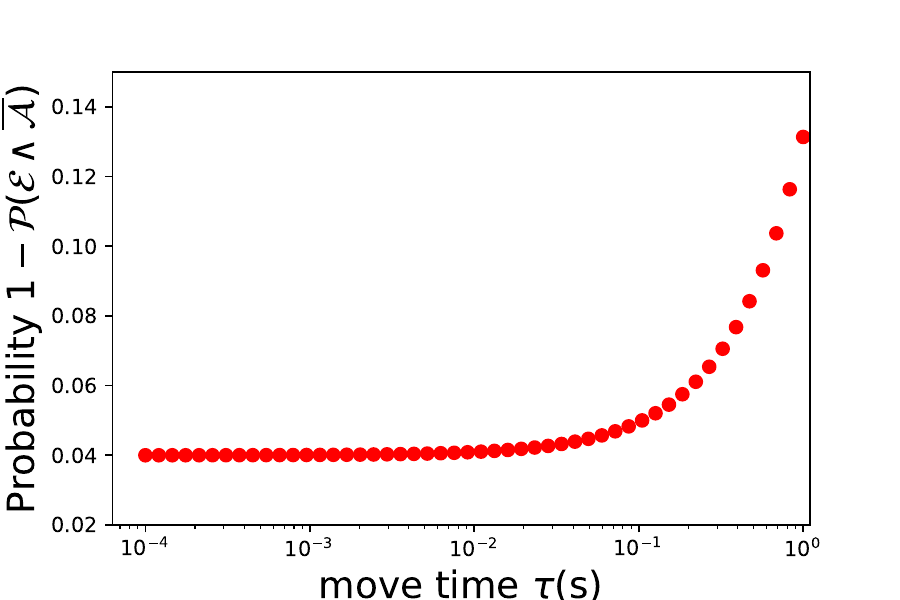}
    \caption{The estimation of the probability of logical qubits being compromised by the CRE when our technique is deployed. In our strategy, the loss percentage drops from $100\%$ to $4\%$ when the logical qubit move time varies between $0.1\mu{}sec$ and $100\mu{}sec$. For long and close to $1s$ move durations, the loss percentage drops from $100\%$ to $15\%$.}
    \label{fig:eval-prob}
\end{figure}

The estimation of this probability ($1 - \mathcal{P}(\mathcal{E} \wedge \mathcal{\overline{A}})$) is illustrated in Fig.~\ref{fig:eval-prob} where $\lambda = 1/10$, and $\tau \in [10^{-4}, 1s]$, $\tau$ in second ($s$).

\section{Discussions and Limitations}
\label{sec:discussions-limitations}

Quantum computers are expected to have a wide range of applications, which is why they generate much interest. However, their alleged Achilles heel of error correction seems vulnerable to a type of error called a cosmic ray event (CRE). Unlike classical errors, CREs produce correlated errors that can destroy the data held in several qubits at the same time. Hardware or distributed systems developed recently to counter this error are limited. Hence, this paper presents a different perspective by developing a hybrid hardware/software-based strategy based on the 2-D surface code with a hardware strategy that limits the phonon propagation radius. The software strategy we propose is to flee the area: move the logical qubits far enough away from the strike's epicenter to maintain our logical information. We provide the necessary specifications on the hardware side to support this approach on the software side easily. We propose a mapping based on surface code that enables easy qubit moves and limits time steps for moving our logical qubit to a safe location regardless of the CRE position. In addition to cosmic rays, ambient radioactivity can also produce correlated errors with high frequency. The ambient radioactivity produces phonons at a rate of about $20$ mHz, while cosmic rays contribute at a rate of $10$ mHz, in a typical $1$ $cm^2$ chip \cite{refId0}. In light of the fact that ambient radioactivity is involved, we can minimize the strike probability (because these particles have low energy) by reducing the cross-section or by adding shielding. Our technique is effective for both cosmic ray phonons and ambient radioactive phonons.

This paper has focused only on superconducting quantum
processors, which require electrons to stay in pairs, and those pairs can be split apart by energy much less than a single eV~\cite{mcewen2021resolving}. Quantum dots and other semiconductor-based approaches are likely susceptible to similar processes but with different constant rates. Ion trap and similar individual-atoms-in-a-vacuum methods (neutral atoms, Rydberg atoms) probably are not susceptible; not only is the probability of an individual atom getting hit vanishingly small but also the mechanism by which the errors could propagate is missing entirely. In an ion trap, an individual atom would get kicked out of the system entirely, and the system would have to recognize that and recover. Quantum error correction can handle qubit loss quite well, but the engineering in keeping the rest of the atoms in place, inserting a new atom where the old one was lost, and rebuilding the code is a lot of work. Ion trap engineers are working on such approaches because their biggest problem is an imperfect vacuum, so stray atoms flying around occasionally collide with their data atoms. Photonic quantum computers certainly would not be susceptible to cosmic rays, except in the necessary photon detectors where their impact is well understood.

However, any scientific work has its limits. As you can see, our solution establishes hardware requirements that cannot yet be met. Hopefully, this work will give hardware engineers a boost of motivation to look in this direction. However, even optimistically, our logical qubit remains vulnerable during the move, and this work only considers one quantum chip. As part of our future work, we will study the combination of this work return and distributed, fault-tolerant protection using the same hybrid hardware-software strategy~\cite{PhysRevLett.129.240502}.

\bibliographystyle{ieeetr}
\bibliography{main}

\begin{thebibliography}{10}

\bibitem{lidar_brun_2013}
D.~A. Lidar and T.~A. Brun, {\em Quantum Error Correction}.
\newblock Cambridge University Press, 2013.

\bibitem{nielsen_chuang_2010}
M.~A. Nielsen and I.~L. Chuang, {\em Quantum Computation and Quantum
  Information: 10th Anniversary Edition}.
\newblock Cambridge University Press, 2010.

\bibitem{gottesman2010introduction}
D.~Gottesman, ``An introduction to quantum error correction and fault-tolerant
  quantum computation,'' in {\em Quantum information science and its
  contributions to mathematics, Proceedings of Symposia in Applied
  Mathematics}, vol.~68, pp.~13--58, 2010.

\bibitem{devitt13:rpp-qec}
S.~J. Devitt, W.~J. Munro, and K.~Nemoto, ``Quantum error correction for
  beginners,'' {\em Reports on Progress in Physics}, vol.~76, no.~7, p.~076001,
  2013.

\bibitem{RevModPhys.87.307}
B.~M. Terhal, ``Quantum error correction for quantum memories,'' {\em Rev. Mod.
  Phys.}, vol.~87, pp.~307--346, Apr 2015.

\bibitem{mcewen2021resolving}
M.~McEwen, L.~Faoro, K.~Arya, A.~Dunsworth, T.~Huang, S.~Kim, B.~Burkett,
  A.~Fowler, F.~Arute, J.~C. Bardin, {\em et~al.}, ``Resolving catastrophic
  error bursts from cosmic rays in large arrays of superconducting qubits,''
  {\em Nature Physics}, vol.~18, pp.~107--111, Dec. 2021.

\bibitem{Wilen_2021}
C.~D. Wilen, S.~Abdullah, N.~A. Kurinsky, C.~Stanford, L.~Cardani,
  G.~D'Imperio, C.~Tomei, L.~Faoro, L.~B. Ioffe, C.~H. Liu, A.~Opremcak, B.~G.
  Christensen, J.~L. DuBois, and R.~McDermott, ``Correlated charge noise and
  relaxation errors in superconducting qubits,'' {\em Nature}, vol.~594,
  pp.~369--373, jun 2021.

\bibitem{Cardani-paper}
L.~Cardani, N.~Casali, G.~Catelani, T.~Charpentier, M.~Clemenza, I.~Colantoni,
  A.~Cruciani, L.~Gironi, L.~Grünhaupt, D.~Gusenkova, F.~Henriques, M.~Lagoin,
  M.~Martinez, S.~Pirro, I.~Pop, C.~Rusconi, A.~Ustinov, F.~Valenti,
  M.~Vignati, and W.~Wernsdorfer, ``Demetra: Suppression of the relaxation
  induced by radioactivity in superconducting qubits,'' {\em Journal of Low
  Temperature Physics}, vol.~199, 04 2020.

\bibitem{Zwanenburgspinqubits}
F.~A. Zwanenburg, A.~S. Dzurak, A.~Morello, M.~Y. Simmons, L.~C.~L. Hollenberg,
  G.~Klimeck, S.~Rogge, S.~N. Coppersmith, and M.~A. Eriksson, ``Silicon
  quantum electronics,'' {\em Rev. Mod. Phys.}, vol.~85, pp.~961--1019, Jul
  2013.

\bibitem{2012majorana}
D.~Rainis and D.~Loss, ``Majorana qubit decoherence by quasiparticle
  poisoning,'' {\em Phys. Rev. B}, vol.~85, p.~174533, May 2012.

\bibitem{martiniss415342021saving}
J.~M. Martinis, ``Saving superconducting quantum processors from decay and
  correlated errors generated by gamma and cosmic rays,'' {\em npj Quantum
  Information}, vol.~7, no.~1, p.~90, 2021.

\bibitem{Clemens2004quantum}
J.~P. Clemens, S.~Siddiqui, and J.~Gea-Banacloche, ``Quantum error correction
  against correlated noise,'' {\em Phys. Rev. A}, vol.~69, p.~062313, Jun 2004.

\bibitem{Nickerson2019analysingcorrelated}
N.~H. Nickerson and B.~J. Brown, ``Analysing correlated noise on the surface
  code using adaptive decoding algorithms,'' {\em {Quantum}}, vol.~3, p.~131,
  Apr. 2019.

\bibitem{van-meter13:_blueprint}
R.~Van{ }Meter and D.~Horsman, ``A blueprint for building a quantum computer,''
  {\em Communications of the ACM}, vol.~53, pp.~84--93, Oct. 2013.

\bibitem{Gidney2021howtofactorbit}
C.~Gidney and M.~Eker{\aa{}}, ``How to factor 2048 bit {RSA} integers in 8
  hours using 20 million noisy qubits,'' {\em {Quantum}}, vol.~5, p.~433, Apr.
  2021.

\bibitem{PhysRevA.76.042319}
J.~Koch, T.~M. Yu, J.~Gambetta, A.~A. Houck, D.~I. Schuster, J.~Majer,
  A.~Blais, M.~H. Devoret, S.~M. Girvin, and R.~J. Schoelkopf,
  ``Charge-insensitive qubit design derived from the cooper pair box,'' {\em
  Phys. Rev. A}, vol.~76, p.~042319, Oct 2007.

\bibitem{krantz2019quantum}
P.~Krantz, M.~Kjaergaard, F.~Yan, T.~P. Orlando, S.~Gustavsson, and W.~D.
  Oliver, ``A quantum engineer's guide to superconducting qubits,'' {\em
  Applied physics reviews}, vol.~6, no.~2, p.~021318, 2019.

\bibitem{PhysRevLett.129.240502}
Q.~Xu, A.~Seif, H.~Yan, N.~Mannucci, B.~O. Sane, R.~Van~Meter, A.~N. Cleland,
  and L.~Jiang, ``Distributed quantum error correction for chip-level
  catastrophic errors,'' {\em Phys. Rev. Lett.}, vol.~129, p.~240502, Dec 2022.

\bibitem{Suzukiburst2022}
Y.~Suzuki, T.~Sugiyama, T.~Arai, W.~Liao, K.~Inoue, and T.~Tanimoto, ``Q3de: A
  fault-tolerant quantum computer architecture for multi-bit burst errors by
  cosmic rays,'' in {\em 2022 55th IEEE/ACM International Symposium on
  Microarchitecture (MICRO)}, pp.~1110--1125, 2022.

\bibitem{patel2017phonon}
U.~Patel, I.~V. Pechenezhskiy, B.~L.~T. Plourde, M.~G. Vavilov, and
  R.~McDermott, ``Phonon-mediated quasiparticle poisoning of superconducting
  microwave resonators,'' {\em Phys. Rev. B}, vol.~96, p.~220501, Dec 2017.

\bibitem{nsanzineza2014trapping}
I.~Nsanzineza and B.~L.~T. Plourde, ``Trapping a single vortex and reducing
  quasiparticles in a superconducting resonator,'' {\em Phys. Rev. Lett.},
  vol.~113, p.~117002, Sep 2014.

\bibitem{henriques2019phonon}
F.~Henriques, F.~Valenti, T.~Charpentier, M.~Lagoin, C.~Gouriou,
  M.~Mart{\'\i}nez, L.~Cardani, M.~Vignati, L.~Gr{\"u}nhaupt, D.~Gusenkova,
  {\em et~al.}, ``Phonon traps reduce the quasiparticle density in
  superconducting circuits,'' {\em Applied physics letters}, vol.~115, no.~21,
  p.~212601, 2019.

\bibitem{raussendorf:PhysRevLett.98.190504}
R.~Raussendorf and J.~Harrington, ``Fault-tolerant quantum computation with
  high threshold in two dimensions,'' {\em Phys. Rev. Lett.}, vol.~98,
  p.~190504, May 2007.

\bibitem{fowler:PhysRevA.86.032324}
A.~G. Fowler, M.~Mariantoni, J.~M. Martinis, and A.~N. Cleland, ``Surface
  codes: Towards practical large-scale quantum computation,'' {\em Phys. Rev.
  A}, vol.~86, p.~032324, Sep 2012.

\bibitem{horsman2012surface-njp}
C.~Horsman, A.~Fowler, S.~Devitt, and R.~Van{ }Meter, ``Surface code quantum
  computing by lattice surgery,'' {\em New Journal of Physics}, vol.~14,
  p.~123011, 2012.

\bibitem{RaussendorfPhysRevLett.98.190504}
R.~Raussendorf and J.~Harrington, ``Fault-tolerant quantum computation with
  high threshold in two dimensions,'' {\em Phys. Rev. Lett.}, vol.~98,
  p.~190504, May 2007.

\bibitem{refId0}
{Cardani, L.}, {Colantoni, I.}, {Cruciani, A.}, {De Dominicis, F.},
  {D\'{}Imperio, G.}, {Laubenstein, M.}, {Mariani, A.}, {Pagnanini, L.},
  {Pirro, S.}, {Tomei, C.}, {Casali, N.}, {Ferroni, F.}, {Frolov, D.}, {Gironi,
  L.}, {Grassellino, A.}, {Junker, M.}, {Kopas, C.}, {Lachman, E.}, {McRae, C.
  R. H.}, {Mutus, J.}, {Nastasi, M.}, {Pappas, D. P.}, {Pilipenko, R.}, {Sisti,
  M.}, {Pettinacci, V.}, {Romanenko, A.}, {Van Zanten, D.}, {Vignati, M.},
  {Withrow, J. D.}, and {Zhelev, N. Z.}, ``Disentangling the sources of
  ionizing radiation in superconducting qubits,'' {\em Eur. Phys. J. C},
  vol.~83, no.~1, p.~94, 2023.

\end{thebibliography}

\end{document}